\documentclass{aa}
\usepackage{graphicx}
\usepackage{multicol}
\usepackage{amsmath}

\usepackage{amsthm}
\usepackage{txfonts}
\usepackage{natbib}
\bibpunct{(}{)}{;}{a}{}{,} 
\usepackage{color}

\newcommand{\fvec}[1]{\mathbf{#1}}
\newcommand{\fdivp}[1]{\fvec{\nabla} \cdot \left(#1\right)}
\newcommand{\fdiv}[1]{\fvec{\nabla} \cdot #1}

\newcommand{\epskin}{\epsilon_{\mathrm{kin}}}
\newcommand{\ndref}[1]{{#1}_{\mathrm{r}}}
\newcommand{\ord}[1]{\mathcal{O}\left({#1}\right)}
\renewcommand{\rho}{\varrho}

\newtheorem{constraint}{Requirement}{\bfseries}{\itshape}

\begin{document}

\title{New numerical solver for flows at various Mach numbers}

\author{F.~Miczek\inst{1}, F.~K.~R{\"o}pke\inst{2,}\inst{3}, \and P.~V.~F.~Edelmann\inst{2,}\inst{3,}\inst{1}}
\authorrunning{F.~Miczek et al.}

\institute{Max-Planck-Institut f\"ur Astrophysik, Karl-Schwarzschild-Str. 1,
  D-85741 Garching, Germany\\
  \email{fab@miczek.de}
  \and Institut f\"ur Theoretische Physik und Astrophysik,
  Universit\"at W\"urzburg, 
  Emil-Fischer-Str. 31,
  D-97074 W\"urzburg, Germany
  \and Heidelberger Institut f{\"u}r Theoretische Studien, Schloss-Wolfsbrunnenweg 35, D-69118 Heidelberg, Germany
}

\date{Received xxxx xx, xxxx / accepted xxxx xx, xxxx}

 
\abstract
{Many problems in stellar astrophysics feature flows at low Mach numbers.
  Conventional compressible hydrodynamics schemes frequently
  used in the field have been developed for the transonic regime and
  exhibit excessive numerical dissipation for these flows.}
{While schemes were proposed that solve hydrodynamics strictly in the
  low Mach regime and thus restrict their applicability, we aim at
  developing a scheme that correctly operates in a wide range of Mach
  numbers.}
{Based on an analysis of the asymptotic behavior of the Euler
  equations in the low Mach limit we propose a novel scheme that is
  able to maintain a low Mach number flow setup while retaining all
  effects of compressibility. This is achieved by a suitable
  modification of the well-known Roe solver.}
{Numerical tests demonstrate the capability of this new scheme to
  reproduce slow flow structures even in moderate numerical
  resolution.}
{Our scheme provides a promising approach to a consistent
  multidimensional hydrodynamical treatment of astrophysical low Mach
  number problems such as
  convection, instabilities, and mixing in stellar evolution.}

\keywords{hydrodynamics -- methods: numerical}

\maketitle
%

\section{Introduction}

Many phenomena in astrophysics can be addressed with hydrodynamical
descriptions. This is because the spatial dimensions of the objects
under consideration are large and the separation between the
scales of interest and the microscopic scales is typically wide enough so that
a hydrodynamical formulation is appropriate\footnote{We note that
  magnetic fields are dynamically relevant in many astrophysical
  processes. Therefore, a description within the framework of
  magnetohydrodynamics is required. This, however, is beyond the scope
  of the work presented here.}.

Astrophysical systems of interest often involve flows in
transonic or supersonic regimes and conventional codes designed for
treating compressible hydrodynamics have been successfully applied in
these situations. A widely used approach to solving hydrodynamics in
astrophysics is based on finite-volume discretization. In particular,
Godunov-like methods are a common tool. Such schemes are naturally
conservative, a highly desirable property in astrophysical
simulations. At the same time, they are also well-suited to be used in 
conjunction with mapped, curvilinear grids
\citep[e.g.,][]{kifonidis2012a, grimm-strele2013a}.

Other astrophysical objects (such as stars) evolve over much
longer times than the dynamical or sound crossing scales, however. Flows that
operate here are highly subsonic, meaning they are characterized by a low Mach
number, which is defined as the ratio of the fluid velocity to the
speed of sound. Flows at a low Mach number challenge the hydrodynamical
treatment in numerical simulations. Conventional finite-volume methods
show difficulties in representing already moderately low Mach numbers
-- depending on the details of the scheme and the numerical resolution,
this is typically observed below Mach numbers of
$10^{-2}$. Higher-order reconstructions may slightly alleviate this problem but are computationally more expensive. In any case, the
success of such approaches is limited to some minimum Mach number.

Therefore, alternatives to the compressible Euler equations are
frequently used in the low-Mach-number regime. Since sound waves are
usually not of interest in the situations to be modeled, it is common
practice to start out from the incompressible version of the equations.
Although this is a useful strategy for terrestrial systems, it
was found to be to be inadequate to describe of many
astrophysical processes. While the gravity source term can be
included in the velocity equation, thermal effects
of expansion or contraction are not accounted for. This makes the
incompressible equations unsuitable for a variety of situations,
including simulations of stellar convection or thermonuclear burning.

For this reason, alternative formulations of hydrodynamics are used
that include thermal effects to some extent. The Boussinesq
approximation \citep[e.g.,][]{sutherland2010a} assumes that all
thermodynamic variables remain very close to a spatially constant
background state and all flows are adiabatic. In astrophysical
situations (e.g., stellar atmospheres), stratification is common so that
the Boussinesq approximation is only acceptable in a very narrow spatial
region.

This limitation is relaxed in the anelastic approximation that allows
for vertical stratifications of the background state. Consequently,
the anelastic equations have been successfully used to study stellar
atmospheres, typically in the context of convective flow phenomena
\citep[e.g.,][]{glatzmaier1984a,miesch2000a,talon2003a,browning2004a}.
But restrictions also apply for such approaches. Most importantly, an
almost adiabatic temperature stratification of a temporarily constant
hydrostatic background state is assumed, which is a fair approximation
in convective but not in radiative regions in stellar interiors. While
approaches exist to overcome this problem \cite[e.g.,][and references
therein]{miesch2005a} and various different anelastic approximations
have been suggested \citep[see][and references therein]{brown2012b},
the application of anelastic schemes to atmospheres with complex
temperature profiles remains challenging.

Another set of modified equations for the regime of low Mach numbers is
implemented in the MAESTRO code \citep{almgren2007a,nonaka2010a}.
This set was specifically derived to filter out the sound waves in the
solution while retaining compressibility effects. This enables the
code to include source terms such as those arising from nuclear burning. The removal of
sound waves also relaxes the CFL constraint on the time steps, which
facilitates covering long simulation times. Despite these
advantages, the validity of the solution is strictly limited to the regime of low
Mach numbers. Therefore, the modeling has to be switched back to the
original Euler equations when the Mach number in some part of the flow
exceeds a certain threshold.

An approach that allows for high and at least moderately-low Mach
numbers (down to $\sim 10^{-3}$) on the same grid is included in the
ANTARES code \citep{muthsam2010a}. It uses the method by
\citet{kwatra2009a} to split the Euler equations into an advective and
an acoustic part. Only the acoustic part is treated with implicit
time-integration, the advective part is integrated explicitly. This
removes the necessity of resolving the sound crossing-time and therefore
makes simulations covering long time-intervals more efficient. It also
improves the accuracy of the solution at low Mach numbers
\citep{happenhofer2013a}.

The MUSIC code \citep{viallet2011a,viallet2013a} uses implicit
time-integration for the complete Euler equations and is thus able to
cover longer timescales than explicit codes. It uses a staggered
grid, which reduces discretization errors for low Mach numbers.

All these approaches either result from modifications of the original
compressible Euler equations, which restricts their applicability to
low Mach numbers, or their ability to represent slow flows is limited
to only moderately low Mach numbers. Here we propose to modify
a numerical solver for compressible hydrodynamics so that it also
correctly reproduces the regime of slow flows. We emphasize that only
the numerical solution technique is altered, but our scheme is still
based on a discretization of the compressible
Euler equations. It is therefore not fundamentally restricted in its applicability to a certain
range in Mach numbers.

Our new asymptotic-preserving scheme has several advantages over other
approaches: It allows simulating situations were fast and slow
flows co-exist in the domain of interest. As we show in this work, its
reduced numerical dissipation considerably improves the reproduction
of small-scale flow features already in moderately slow
flows. Moreover, the additional effort in implementing the method into
existing frameworks of compressible hydrodynamics solvers is
modest. The result is a versatile tool for astrophysical fluid
dynamics that can be used for example to simulate dynamical phases of
stellar evolution and the effect of instabilities on mixing and energy
transport in stars.

The technique we employ in our scheme is a preconditioning of the
  numerical flux function. By modifying the spatial discretization of
the Euler equations, it considerably improves the numerical
representation of flows at low Mach numbers independently of the
particular choice of time discretization.  Preconditioning methods
have been extensively discussed in the mathematical literature
\citep[for a review see][]{turkel1999a}, but often with emphasis on
the solution of steady-state problems. These methods are employed in
various contexts in engineering, such as tunnel fires
\citep{birken2008a}, gas turbines \citep{renze2008a}, and airfoils
\citep{nemec2000a}. For unsteady flows at all Mach numbers,
  \citet{mary2000a} constructed a scheme based on a dual time-stepping
  technique, which involves solving a preconditioned
  steady-state problem at every time step. Compared to their scheme,
  the approach presented here merely modifies the numerical
  fluxes and needs no additional solution steps. It is
  therefore simpler to implement in existing simulation codes.

  To our knowledge, preconditioning techniques for flows at low Mach numbers
have not yet been applied to astrophysical problems. We therefore give
an extensive introduction to the field by rewriting the Euler
equations of fluid dynamics in nondimensional form in
Sect.~\ref{sect:nondim} and analyzing their solution space in the limit of low
Mach numbers in Sect.~\ref{sect:lowmach}. Based on this analysis,
we formulate requirements for a consistent low-Mach-number
hydrodynamics solver. To illustrate the difficulties numerical schemes
encounter for slow flows and the solution we propose, we introduce a
version of the Gresho vortex problem in Sect.~\ref{sect:gresho}. In
Sects.~\ref{sect:discr} and \ref{sect:num_flux} we discuss the spatial
discretization of the Euler equations in finite-volume approaches and
numerical flux functions. This extensive prelude allows
us to analyze the scaling behavior of numerical solutions in the limit of low
Mach numbers in Sect.~\ref{sect:low-mach-scaling}, where we
discuss the reason why conventional solvers fail for slow flows. In
Sect.~\ref{sect:preconditioner} we introduce preconditioning
techniques to overcome this problem and propose a new flux
preconditioner that ensures correct scaling at low Mach
numbers. Numerical tests demonstrating the success of our new scheme
are presented in Sect.~\ref{sect:tests}. In Sect.~\ref{sect:concl} we
draw conclusions based on the requirements on low-Mach-number
hydrodynamics solvers set out in Sect.~\ref{sect:lowmach}.

\section{Nondimensional Euler equations}
\label{sect:nondim}

While the scales of astrophysical objects are huge and many
processes involve rather high velocities, the viscosities of astrophysical
media are not extraordinarily large. Therefore, the Reynolds numbers
characterizing astrophysical flows are high and it is justified
to treat matter as an ideal fluid. The motion of an ideal fluid is
described by the Euler equations,
\begin{equation}\label{eq:euler}
  \frac{\partial \fvec{U}}{\partial t} +
  \frac{\partial \fvec{F}_x}{\partial x} +
  \frac{\partial \fvec{F}_y}{\partial y} +
  \frac{\partial \fvec{F}_z}{\partial z} = \fvec{S},
\end{equation}
where the state vector of conservative variables,
\begin{equation}
  \label{eq:def-U}
  \fvec{U} = \left(
      \rho,
      \rho u,
      \rho v,
      \rho w,
      \rho E,
      \rho X
  \right)^\top,
\end{equation}
is composed of the mass density $\rho$, the Cartesian components $u$, $v$,
and $w$ of the velocity $\fvec{q}$, and the specific total (internal plus
kinetic) energy $E$ of the fluid. Optionally, some scalar values $X$
can be introduced, representing composition, for instance. The vector
$\fvec{S}$ on the right-hand side of Eq.~\eqref{eq:euler} contains
source terms due to external forces or
reactions, for example.

The speed of sound is given by the change of pressure with mass density in
an isentropic process (entropy $S=\mathrm{const}$):
\begin{equation*}
  c=\sqrt{\left.\frac{\partial p}{\partial\rho}\right|_S}.
\end{equation*}
The ratio of the fluid velocity $q \equiv \sqrt{u^2+v^2+w^2}$ to the
speed of sound defines the Mach number,
\begin{equation*}
M \equiv \frac{q}{c}.
\end{equation*} 

To make the treatment independent of the choice of a specific system of
units, we rewrite the quantities in nondimensional form (except for
$X$, which, for our purposes, is usually unit-free). At the
same time, this form provides a clear picture of the number of
parameters the setup depends on. Nondimensionalization of the
equations is achieved by decomposing every quantity into a reference
value (marked with subscript ``$\mathrm{r}$'' in the following, e.g.,\
$\ndref{\rho}$) and a dimensionless number (e.g., $\hat{\rho}$). The
reference value is chosen such that the nondimensional quantity has a
value of order unity. A suitable system of reference quantities is a
reference mass density $\ndref{\rho}$, a reference fluid velocity
$\ndref{q}$, a reference speed of sound $\ndref{c}$, and a reference
length scale $\ndref{x}$. Other quantities, such as
reference time, reference pressure, and reference specific total energy, can be derived
from this set:
\begin{equation*}
  \ndref{t} = \frac{\ndref{x}}{\ndref{u}},
  \quad
  \ndref{p} = \ndref{\rho} \ndref{c}^2,
  \quad
  \ndref{E} = \ndref{e} = \ndref{c}^2.
\end{equation*}
The chosen form of reference pressure results from the expression for
the speed of sound in an ideal gas, $c^2=\gamma\, p/\rho$, where
$\gamma$ denotes the adiabatic index.  The reference value of the
specific internal energy $\ndref{e}$ is set following the ideal gas law,
$p = (\gamma - 1) \rho e$. The same reference value is also used for
the specific total energy $\ndref{E}$. This is an appropriate choice
in the regime of low Mach numbers, where the contribution of kinetic
energy is negligible. We note that although we argue with expressions
for the ideal gas, the strategy of nondimensionalization used
here applies more generally since the specific choices of reference
quantities are somewhat arbitrary.

In nondimensional form, the flux vectors of Eq.~\eqref{eq:euler}
corresponding to the conservative variables \eqref{eq:def-U} read
\begin{eqnarray}\label{eq:cart_fluxes}
  \hat{\fvec{F}}_x \;\;  = \;\;  \left(
    \begin{array}{c}
      \hat\rho \hat u \\
      \hat\rho \hat u \hat u + \frac{\hat p}{\ndref{M}^2}\\
      \hat\rho \hat u \hat v \\
      \hat\rho \hat u \hat w \\
      \hat\rho \hat u \hat E + \hat u \hat p \\
      \hat\rho \hat u  X
    \end{array}
  \right), \qquad
  \hat{\fvec{F}}_y &=& \left(
    \begin{array}{c}
      \hat\rho \hat v \\
      \hat\rho \hat v \hat u \\
      \hat\rho \hat v \hat v + \frac{\hat p}{\ndref{M}^2}\\
      \hat\rho \hat v \hat w \\
      \hat\rho \hat v \hat E + \hat v \hat p \\
      \hat\rho \hat v  X
    \end{array}
  \right),\nonumber\\
  \hat{\fvec{F}}_z &=& \left(
    \begin{array}{c}
      \hat\rho \hat w \\
      \hat\rho \hat w \hat u \\
      \hat\rho \hat w \hat v \\
      \hat\rho \hat w \hat w + \frac{\hat p}{\ndref{M}^2}\\
      \hat\rho \hat w \hat E + \hat w \hat p\\
      \hat\rho \hat w X
    \end{array}
  \right).
\end{eqnarray}
They involve pressure $\hat p$ that is related to the conservative
variables by an equation of state.

These Euler equations express continuity of mass, balance
of the components of momentum, balance of energy and, optionally,
balance of scalar values. Note that the homogeneous Euler equations
(i.e., $\fvec{S}=\fvec{0}$) depend on a single, nondimensional reference
quantity: the reference Mach number $\ndref{M} = \ndref{q}/\ndref{c}$.
In the following, all quantities are nondimensional, and we drop the hat
over the nondimensional quantities for convenience.

\section{Fluid dynamics at low Mach numbers}
\label{sect:lowmach}

In the limit of $\mathit{M} \to 0$, the Euler equations
(\ref{eq:euler}) approach the incompressible equations of fluid
dynamics, that is, the substantial derivative of the mass density, $\partial
\rho / \partial t + \fvec{q} \cdot \nabla \rho$, vanishes. To show this, we follow the
approach taken by \citet{guillard1999a} and expand the nondimensional
quantities in terms of the reference Mach number,
\begin{equation}\label{eq:exp}
  \begin{array}{rclclclcl}
    {\rho} &=& {\rho}_0 &+& {\rho}_1 \ndref{M} &+&
    {\rho}_2 \ndref{M}^2     &+& \mathcal{O}\left(\ndref{M}^3\right),\\
    {\fvec{q}} &=& {\fvec{q}}_0 &+& {\fvec{q}}_1 \ndref{M} &+&
    {\fvec{q}}_2 \ndref{M}^2 &+& \mathcal{O}\left(\ndref{M}^3\right),\\
    {p} &=& {p}_0 &+& {p}_1 \ndref{M} &+&
    {p}_2 \ndref{M}^2        &+& \mathcal{O}\left(\ndref{M}^3\right),\\
    {E} &=& {E}_0 &+& {E}_1 \ndref{M} &+& {E}_2 \ndref{M}^2 &+&
    \mathcal{O}\left(\ndref{M}^3\right).
  \end{array}
\end{equation}
Each term in the expansions may be a function of space and time.  For
simplicity, we consider the homogeneous Euler equations ($\fvec{S} =
\fvec{0}$) in the following. With the above expressions, they read
\begin{align}
    \frac{\partial}{\partial {t}} \, {\rho}_0 +
    \fdivp{ {\rho}_0 {\fvec{q}}_0  } + \mathcal{O}(\ndref{M}) = 0,\label{eq:cont}\\
     \begin{aligned}\frac{\partial}{\partial {t}} \left({\rho}_0
      {\fvec{q}}_0\right) + \fdivp{{\rho}_0 {\fvec{q}}_0
      \otimes {\fvec{q}}_0 } + &\\ \frac{1}{\ndref{M}^2}
    \fvec{\nabla}{p}_0 + \frac{1}{\ndref{M}} \fvec{\nabla}{p}_1 +
    \fvec{\nabla}{p}_2 & +  \mathcal{O}(\ndref{M}) = 0, \end{aligned}\label{eq:mom}\\
    \frac{\partial}{\partial {t}} \left({\rho}_0 {E}_0
    \right) + \fdivp{ {\rho}_0 {E}_0 {\fvec{q}}_0 + {p}_0
      {\fvec{q}}_0 } + \mathcal{O}(\ndref{M}) = 0. \label{eq:en}
  \end{align}

Obviously, terms in these equations scale differently with Mach number. To reach
convergence in the limit of $\mathit{M} \to 0$, it is necessary that the
expressions of equal order in Mach number on the left-hand side vanish
individually for the non-positive powers of $\ndref{M}$. We therefore find two simple relations,
\begin{equation} \label{eq:const_p}
  \fvec{\nabla}{p}_0 = \fvec{\nabla}{p}_1 = 0,
\end{equation}
from the terms of order $\ndref{M}^{-2}$ and $\ndref{M}^{-1}$
in the momentum equations (\ref{eq:mom}), respectively. Thus, pressure must be
constant in space up to fluctuations of order $\ndref{M}^2$ \citep{guillard1999a}:
\begin{equation}\label{eq:p_scaling_incompr}
  {p} \left( {\fvec{x}},{t} \right) = 
  {p}_0({t}) + {p}_2( {\fvec{x}},{t}) \, \ndref{M}^2 + \mathcal{O}(\ndref{M}^3).
\end{equation}
Time variations on the spatially constant ${p}_0$ can only be
imposed through source terms, which we
set to zero for the considerations in this article, or by the boundary conditions. In case of
open boundary conditions, ${p}_0$ is set by the exterior pressure, which
we assume to be temporally fixed following \citet{guillard1999a}. Thus,
${p}_0$ is constant in space and time \citep[see also][]{dellacherie2010a}.

The expansion in Mach number is also applied to the equation of
state. For simplicity, an ideal gas is assumed here. Inserting the
expansion of the nondimensional quantities (\ref{eq:exp}) into the
nondimensionalized equation of state,
\begin{equation*}
  {p} = (\gamma-1)\left[ {\rho} {E} - \frac{\ndref{M}^2}{2}{\rho}{q^2}\right],
\end{equation*}
we obtain
\begin{equation} \label{eq:eos} {p}_0=(\gamma-1){\rho}_0
  {E}_0 + \mathcal{O}(\ndref{M}).
\end{equation}
With Eqs.~(\ref{eq:const_p}) and (\ref{eq:eos}), the energy equation
(\ref{eq:en}) can be transformed into
\begin{equation*}
  \frac{\partial {p}_0}{\partial{t}} + \gamma
   {p}_0\, \fdiv{{\fvec{q}}_0}  + \mathcal{O}(\ndref{M})= 0.
\end{equation*}
As argued above, ${p}_0$ is constant in time and thus the velocity
field has to be divergence-free in zero Mach number limit,
\begin{equation}  \label{incomp-div}
  \fdiv{\fvec{{q}}_0}=0,
\end{equation}
recovering the incompressible flow when inserted into
Eq.~\eqref{eq:cont}.  We emphasize that according to
Eq.~(\ref{eq:p_scaling_incompr}), pressure fluctuations within the
compressible Euler equations scale with the square of the
reference Mach number as the solution convergences to a solution of
the incompressible equations.

At the same time, however, stationary fluids with no mean velocity
(i.e., at zero Mach number) can support the propagation of sound waves
with arbitrarily small velocity fluctuations. We perform a standard
linear stability analysis \citep[e.g.,][]{landaulifshitz6eng}: For a
fluid at rest with uniform mass density $\rho_0 = \text{const}$ and
pressure $p_0= \text{const}$, deviations from this state can be
expressed in terms of a small Mach number ($\ndref{M}\ll 1$) following
the expressions (\ref{eq:exp}), except that in this case $\rho_0$ and
$p_0$ are independent of space and time. As before, we insert these
expressions into the Euler equations (\ref{eq:euler}), but, in
contrast to the analysis of the incompressible limit, we now keep
terms up to order $\ndref{M}$. This results in
\begin{equation*}
  \frac{\partial{\rho}_1}{\partial{t}} + 
  {\rho}_0\,
  \fdiv{{\fvec{q}}_1} + \mathcal{O}(\ndref{M})
  = 0
\end{equation*}
for the continuity equation and 
\begin{equation*}
  {\rho}_0
  \frac{\partial{\fvec{q}}_1}{\partial{t}}
  + \frac{1}{\ndref{M}^2}
  \fvec{\nabla} {p}_1  + \mathcal{O}(\ndref{M})
  = 0
\end{equation*}
for the momentum equation. Instead of invoking the
energy equation, we assume that pressure fluctuations result from
reversible, adiabatic processes and thus express them in terms of mass
density fluctuations using the nondimensional speed of sound:
\begin{equation*}
  {p}_1 =
  \left.\frac{\partial{p}}{\partial{\rho}}\right|_{S}
  \cdot {\rho}_1
  = {c}^2 \cdot {\rho}_1.
\end{equation*}
This transforms the continuity equation (dropping terms of order
$\ndref{M}$) into
\begin{equation}\label{eq:contin_sound}
  \frac{\partial{p}_1}{\partial{t}} + 
  {\rho}_0 {c}^2 \,
  \fdiv{{\fvec{q}}_1}
  = 0.
\end{equation}
Together with the momentum equations, we obtain a closed system for
the evolution of ${\fvec{q}}_1$ and ${p}_1$. Introducing the velocity
potential $\psi$,
\begin{equation*}
\fvec{q}_1 \equiv\nabla  \psi,
\end{equation*}
we arrive at the linear wave equation 
\begin{equation*}
\frac{\partial^2 {\psi}}{\partial {t}^2} +
\frac{{c}^2}{\ndref{M}^2} \, \Delta {\psi} = 0
\end{equation*}
describing sound waves propagating at speed ${c}/\ndref{M}$. This
derivation demonstrates that the Euler equations permit sound waves
with arbitrarily small velocity fluctuations as solutions at
arbitrarily small reference Mach numbers.

This result has important implications for the solutions of the
compressible Euler equations at low Mach numbers: For sound waves,
pressure fluctuations are on the same order in $\ndref{M}$ as velocity
fluctuations, as can be seen in Eq.~\eqref{eq:contin_sound}. Thus, in
contrast to the solutions approaching the incompressible regime
discussed above, pressure fluctuations scale \emph{linearly} with the
reference Mach number, that is,
\begin{equation*}
p(\fvec{x}, t) = p_0 + p_1(\fvec{x},t) \, \ndref{M} + \mathcal{O}(\ndref{M}^2).
\end{equation*}
Sound waves are no solutions of the incompressible
equations. Consequently, the compressible Euler equations permit two
distinct classes of solutions in the limit of low Mach numbers:
incompressible flows and sound waves. \citet{schochet1994a} and
\citet{dellacherie2010a} proved that these types of solution decouple
as the Mach number of the flow decreases.

This implies an important requirement for correct solutions of the
compressible Euler equations in the regime of low Mach numbers:

\begin{constraint}
  \label{constraint_1}
  If the initial condition for the compressible, homogeneous Euler
  equations is chosen to be an incompressible flow (i.e., it is
  `well-prepared' such that pressure fluctuations scale with
  $\ndref{M}^2$), then the solution must stay in the incompressible
  regime.
\end{constraint}

This requirement holds for continuous solutions.
At the same time, it is an important quality criterion for numerical
schemes not to violate this constraint. For this, an appropriate
discretization is necessary, as we discuss in
Sects.~\ref{sect:low-mach-scaling} and \ref{sect:preconditioner}.

Another constraint results from considering the evolution of kinetic
energy. Its density, $ \epskin = \frac{1}{2} \rho q^2 $,
is usually treated as part of the total energy $\rho E$ in the Euler
equations. However, a separate evolution equation for the kinetic
energy can be derived from Eq.~\eqref{eq:euler} in the case of vanishing
source terms in the equations of mass and momentum conservation:
\begin{equation*}
  \frac{\partial\epskin}{\partial t} +
  \fdiv{\left[ \left(\epskin+\frac{p}{\ndref{M}^2}\right) \fvec{q} \right]}
  = \frac{p}{\ndref{M}^2}\ \fvec{\nabla}\cdot\fvec{q}.
\end{equation*}
This equation is in strong conservation form, except for the source
term on the right-hand side. However, since the velocity divergence
vanishes for incompressible flows, we identify another requirement for
the correct solution in the low Mach number regime:
\begin{constraint}
  \label{constraint_2}
  For solutions to the compressible, homogeneous Euler equations in the limit of low Mach
  numbers, the total kinetic energy is conserved.
\end{constraint}
\noindent This implies that no dissipation takes place and establishes
another quality criterion for numerical discretization.

The two requirements pose severe challenges to a numerical treatment
of flows at low Mach numbers with conventional Godunov-type solvers. Such
approaches cannot guarantee the correct asymptotic behavior because
their solution strategy conflicts with the well-preparedness of the
initial condition in each time step.  The idea of Godunov-like schemes
is to advance the hydrodynamic states by discretizing the conserved
quantities in a way that discontinuities appear at the cell
interfaces. This defines Riemann problems for which an analytic
solution is known (in practical implementations, approximate solutions
are constructed) so that the hydrodynamical fluxes over the cell edges
can be determined. The introduction of discontinuities and the wave
pattern resulting from the solution to the Riemann problem, however,
is incompatible with flows at low Mach numbers near the incompressible
regime. This is illustrated and further discussed in
Sect.~\ref{sect:low-mach-scaling}.  In addition to this fundamental
problem, we encounter a practical problem for the numerical
discretization. In the limit of low Mach numbers, the fluid velocity is
orders of magnitude lower than the sound speed. This disparity of scales leads
to a stiffness of the system of equations to be solved.

Although requirements~\ref{constraint_1} and \ref{constraint_2}
are not easy to fulfill in the discretized version, it still appears
worthwhile to begin from the compressible equations of fluid
dynamics. In contrast to other methods, no a priori approximations
enter the basic equations. Therefore, schemes based on this approach 
promise to be applicable in flow regimes covering a wide
range of Mach numbers, from incompressible flows over moderate Mach
numbers to the supersonic regime. In the following, we examine the
behavior of a widely used Godunov-like scheme based on the solver
suggested by \citet{roe1981a}. We propose a modification to this
scheme, which ensures correct low-Mach-number scaling and also
alleviates the stiffness problems. The appeal of this approach
is that it retains the advantages of conventional Godunov-type solvers
of the compressible equations of hydrodynamics, such as flexibility of
application and high efficiency of numerical solvers, while allowing
for modeling flow regimes that were inaccessible without the proposed
modifications.

\section{Testbed: the Gresho vortex problem}
\label{sect:gresho}

Before discussing the properties of numerical hydrodynamics solvers in
more detail, we set out a test problem that illustrates the problems
of conventional schemes and allows us to demonstrate the advantages of
the proposed modifications. A suitable setup is a rotating flow in
form of the Gresho vortex \citep{gresho1990a}. This vortex is a
time-independent solution of the incompressible, homogeneous Euler equations where
centrifugal forces are exactly balanced by pressure gradients. It was
first applied to the compressible Euler equations by
\cite{liska2003a}.

The problem is set up in a two-dimensional box. For
nondimensionalization, we choose the reference length $\ndref{x}$ to equal to
domain size, so that the spatial coordinates $x$ and $y$ extend over $[0,1]$ with periodic boundary
conditions. This domain is filled with an ideal gas of constant mass
density, which sets our reference mass density,
that is, $\rho(x,y)=\textrm{const} =\ndref{\rho}$. 

The rotation of the vortex,
which is centered at $(x,y) = (0.5,0.5)$, is initiated by imposing a simple
angular velocity distribution of
\begin{equation*} 
  u_{\phi} (r) = 
  \begin{cases}
    5 r, & 0 \le r < 0.2 \\
    2-5 r, & 0.2 \le r < 0.4 \\
    0, & 0.4 \le r,
  \end{cases}
\end{equation*}
given in units of the reference fluid velocity $\ndref{q}$. Here,
$r=\sqrt{\left(x-0.5\right)^2+\left(y-0.5\right)^2}$ denotes the
radial coordinate of the rotating flow. The velocity reaches its
maximum value $u_{\phi,\mathrm{max}}=1$ at
$r=0.2$. For convenience, we arrange the reference time, $\ndref{t}$, to equal the
period of the vortex. This is achieved by setting the reference
velocity to 
\begin{equation*}
\ndref{q} = \frac{0.4\pi \ndref{x}}{\ndref{t}}. 
\end{equation*}

To stabilize the flow, we set the pressure such that its gradient
gives rise to the required centripetal force:
\begin{equation} \label{eq:gresho_pressure}
  p(r)  = \begin{cases}
    p_0 + \frac{25}{2} r^2, & 0 \le r < 0.2 \\
    p_0 + \frac{25}{2} r^2 + 4 \cdot (1-5 r-\ln 0.2 + \ln r), & 0.2
    \le r < 0.4 \\
    p_0 - 2 + 4 \ln 2, & 0.4 \le r.
  \end{cases}
\end{equation}
This pressure profile is arranged to be continuous and differentiable
everywhere and has a free constant $p_0$. We use this parameter to
conveniently adjust the maximum Mach number of the problem. The Mach
number 
\begin{equation*}
M (r) = \frac{u_{\phi}(r)}{\sqrt{\gamma p(r)/\rho_0}}
\end{equation*}
reaches its maximum, $M_{\mathrm{max}}$, at $r=0.2$. Depending on the
intended maximum Mach number of the vortex flow, we set the
pressure parameter to
\begin{equation}
\label{eq:p0}
  p_0 = \frac{\rho_0\, u_{\phi,\mathrm{max}}^2}{\gamma
    M_{\mathrm{max}}^2} - \frac{1}{2}.
\end{equation}
We note that setting the reference Mach number $\ndref{M} =
M_{\mathrm{max}}$ implies that $\ndref{p} = p_0$.

When choosing $p_0$ as in Eq.~\eqref{eq:p0}, it follows from Eq.~(\ref{eq:gresho_pressure}) that
pressure fluctuations scale with the square of the Mach number,
\begin{equation*}
\frac{p - p_0}{p_0} \propto M^2_{\mathrm{max}}.
\end{equation*}
Thus, according to the discussion in Sect.~\ref{sect:lowmach}, the
initial condition is well prepared, meaning that the flow
is in the incompressible regime and no sound waves occur. The Gresho
vortex as set up here is a stationary solution to the incompressible
Euler equations and thus should not change with time. Any deviations
from stationarity observed in the evolution of the Gresho vortex
can be attributed to discretization errors of the applied numerical scheme. We therefore use this
problem to check the quality of hydro solvers in the regime of
low Mach numbers.

We emphasize that although the Gresho vortex problem is a simplistic
setup, it still provides a relevant test for realistic
applications of numerical hydro-solvers. While discretization
errors can usually be alleviated with higher spatial resolution,
complex flows as encountered in many astrophysical
situations still pose problems. In particular for turbulent flows,
a higher resolution of the domain will usually reveal smaller-scale
structures near the cutoff at the grid scale. These are again
poorly resolved, and discretization errors are most pronounced
here. Therefore, the quality of a numerical scheme for modeling flows at low
Mach numbers reflects this in its ability to adequately represent
such structures. In our Gresho vortex testbed, we therefore
intentionally chose a discretization of the domain in only
$40\times40$ grid cells.

\begin{figure}
\includegraphics[width=\linewidth]{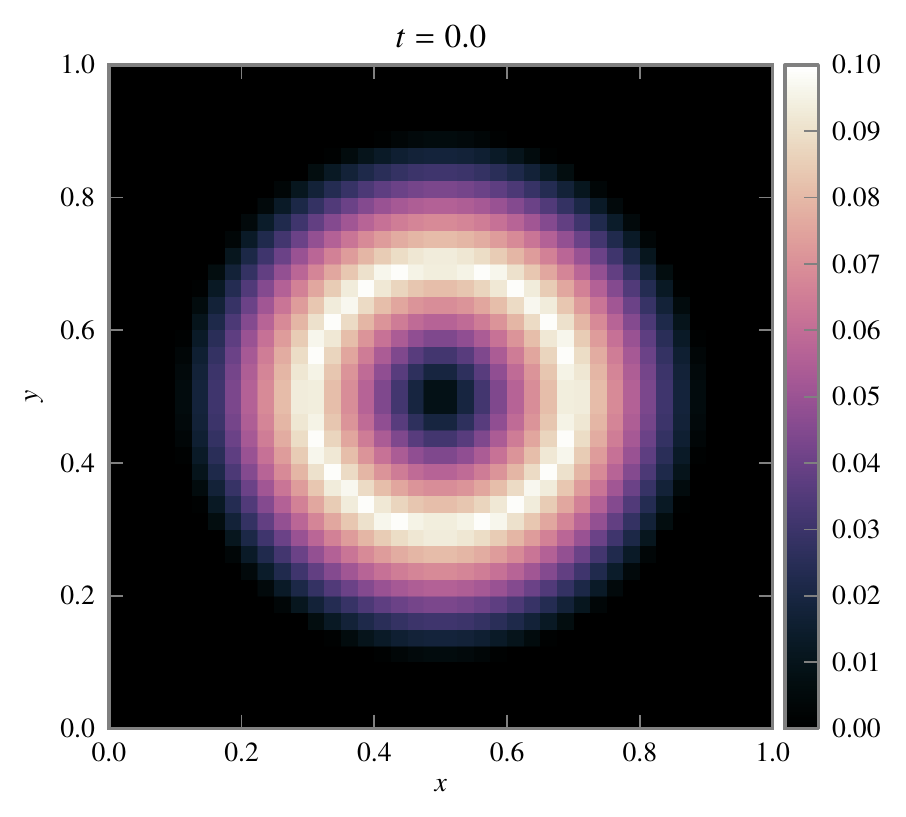}
\caption{Setup of the Gresho vortex problem for $M_{\mathrm{max}} =
  0.1$. The Mach number is color coded.}
\label{fig:gresho_setup}
\end{figure}

The initial setup of the Gresho vortex outlined above is shown in
Fig.~\ref{fig:gresho_setup}. It is advanced in time from $t=0.0$ to
$t=1.0$, corresponding to one full revolution of the vortex. The
choice of the time discretization method has only a minor effect on the
results of our tests. As a standard for the examples shown below, we
chose the implicit, third-order Runge--Kutta method ESDIRK34
\citep{kennedy2001a}.  Because of the integrated error estimator of this
method, adaptive time-step control might be used in
principle. However, we instead employed an advective CFL criterion to determine the time step size $\Delta t$, 
\begin{equation}
\label{eq:cfl_adv}
\Delta t = \frac{c_\mathrm{CFL}}{N_\mathrm{dim}\min\left(\dfrac{\Delta}{|q_n|}\right)},
\end{equation}
where the CFL factor is set to $c_\mathrm{CFL} = 0.5$, and the number
of spatial dimensions of our problem is $N_\mathrm{dim} = 2$. This
implicit method is particularly well suited for our purposes. At low
Mach numbers, its numerical efficiency is expected to be higher than
that of explicit schemes. Moreover, it allows evolving the problem at
various Mach numbers over a given physical time with the same number
of numerical time steps. This increases the comparability between the
runs because potential numerical dissipation due to temporal
discretization is kept constant.

However, to substantiate our claim that the success of our new method
for simulating flows at low Mach numbers be independent of the employed
temporal discretization, we also ran some test problems with the
time-explicit RK3 scheme of \citet{shu1988a}.  Implementation details
and efficiency tests of the temporal discretization will be presented
in a forthcoming publication.

\section{Spatial discretization}
\label{sect:discr}

Our approach to modeling flows at low Mach numbers is based on the
compressible Euler equations and uses standard discretization
methods. Here, we discuss the spatial discretization of the equations
by following the \emph{method of lines} \citep[e.g.,][]{toro2009a} to
arrive at a semi-discrete scheme. We perform a
finite-volume discretization starting from the Euler equations \eqref{eq:euler},
\begin{equation} \label{eq:euler_curvi_1}
 \frac{\partial \fvec{U}}{\partial t} + 
  \fvec{\nabla} \cdot \fvec{\mathcal{F}}
  =\fvec{S}
  \hspace{2em}
  \mathrm{with}
  \hspace{2em}
  \mathcal{F} = \left( 
    \begin{array}{c}
      \fvec{F}_x\\
      \fvec{F}_y\\
      \fvec{F}_z\\
    \end{array}
  \right).
\end{equation}  
For simplicity and clarity of notation, we use a general flux
vector~$\fvec{F}$ instead of writing out its Cartesian components in the
following. This general form can be expressed in terms of the normal velocity~$\fvec{q}_n$,
its magnitude~$q_n$, and its normalized, Cartesian components~$n_x$,
$n_y$, and $n_z$:
\begin{equation}
  \fvec{F} = \begin{pmatrix}
    \rho q_n\\
    \rho u q_n + n_x \frac{p}{\ndref{M}^2}\\
    \rho v q_n + n_y \frac{p}{\ndref{M}^2}\\
    \rho w q_n + n_z \frac{p}{\ndref{M}^2}\\
    (\rho E + p) q_n
  \end{pmatrix}.
  \label{eq:flux-general}
\end{equation}
For instance, for the flux in $x$-direction of Cartesian coordinates
we have $\fvec{q}_n = (u,0,0)^\top$, $q_n = u$ and $n_x =1$, $n_y =
0$, $n_z=0$. This results in the expression for $\fvec{F}_x$ as in Eq.
(\ref{eq:cart_fluxes}).

Spatial discretization is performed in Cartesian coordinates by
partitioning the domain of interest into a regular, equidistant grid
of $N_x \times N_y \times N_z$ cells.  This has the advantage that the
cell widths are uniform, that is,  $\Delta x = \Delta y = \Delta z =
\Delta$, simplifying some of the further expressions. However, an
equidistant grid is not required, and the expressions
given below can
be easily generalized. Integer values
of the coordinates refer to cell centers, for instance, $(i,j,k)$. Cell faces
are denoted by half integer values in the corresponding direction. For
example, the interface between cell $i$ and cell $i+1$ in the
$x$-direction is marked with $i+1/2$.

Integration of (\ref{eq:euler_curvi_1}) over the volume
$\Omega_{i,j,k}$ of a cell $(i,j,k)$ gives
\begin{equation*}
  \int_{\Omega_{i,j,k}} \frac{\partial \fvec{U}}{\partial t} \mathrm{d}\Omega + 
  \int_{\Omega_{i,j,k}}
  \fvec{\nabla} \cdot \fvec{\mathcal{F}}\ 
  \mathrm{d}\Omega
  = \int_{\Omega_{i,j,k}} \fvec{S}\ \mathrm{d}\Omega.
\end{equation*}
The first term on the left-hand side is the time derivative of the
integral of the conserved variables over a cell of volume
$V_{i,j,k}$. We thus introduce cell-averaged quantities:
\begin{equation*}
  \fvec{U}_{i,j,k} = \frac{1}{V_{i,j,k}}
  \int_{\Omega_{i,j,k}} \fvec{U}\ d\Omega.
\end{equation*}
The source term on the right-hand side of the equation is treated
analogously, and we arrive at
\begin{equation*}
  \frac{\partial\fvec{U}_{i,j,k}}{\partial t} + 
  \frac{1}{V_{i,j,k}}
  \oint_{\partial\Omega}
  \fvec{\mathcal{F}} \cdot \fvec{n}\ d\mathcal{S}
  = \fvec{S}_{i,j,k}.
\end{equation*}
The surface integral is discretized by defining \emph{numerical
  fluxes} from the integrated fluxes through the six surfaces of the
three dimensional cell $(i,j,k)$:
\begin{equation}
\label{eq:num_fluxes}
  \begin{array}{rcl}
    \fvec{F}_{i+1/2,j,k} &=& \oint_{\partial\Omega(i+1/2,j,k)}
    \fvec{F}_x\ d\mathcal{S}, \\ 
    \fvec{F}_{i-1/2,j,k} &=& \oint_{\partial\Omega(i-1/2,j,k)}
    \fvec{F}_x\ d\mathcal{S}, \\[2ex]
    \fvec{F}_{i,j+1/2,k} &=& \oint_{\partial\Omega(i,j+1/2,k)}
    \fvec{F}_y\ d\mathcal{S}, \\ 
    \fvec{F}_{i,j-1/2,k} &=& \oint_{\partial\Omega(i,j-1/2,k)}
    \fvec{F}_y\ d\mathcal{S}, \\[2ex]
    \fvec{F}_{i,j,k+1/2} &=& \oint_{\partial\Omega(i,j,k+1/2)}
    \fvec{F}_z\ d\mathcal{S}, \\ 
    \fvec{F}_{i,j,k-1/2} &=& \oint_{\partial\Omega(i,j,k-1/2)}
    \fvec{F}_z\ d\mathcal{S}.
  \end{array}
\end{equation}
This results in the semi-discrete version of the Euler equations,
\begin{equation}
  \label{eq:finitevolume-disc}
  \begin{array}{rll}
    \frac{\partial}{\partial t} \fvec{U}_{i,j,k} + 
    (V_{i,j,k})^{-1}
    \big(
      & \fvec{F}_{i+1/2,j,k} - \fvec{F}_{i-1/2,j,k} \, + & \\
      & \fvec{F}_{i,j+1/2,k} - \fvec{F}_{i,j-1/2,k} \, + &\\
      & \fvec{F}_{i,j,k+1/2} - \fvec{F}_{i,j,k-1/2} &
     \big)
    = \fvec{S}_{i,j,k}.
  \end{array}
\end{equation}
At the domain edges, boundary conditions must be implemented either by
introducing layers of ghost cells or by local modification of the
discretization scheme \citep[see, e.g.,][]{toro2009a}.

\section{Numerical flux functions}
\label{sect:num_flux}

The key ingredient in any finite-volume scheme is constructing
the numerical fluxes defined in Eq.~\eqref{eq:num_fluxes}.  The
discretization described above stores cell-averaged quantities at the
cell centers. According to Eq.~\eqref{eq:finitevolume-disc}, these
quantities are evolved in time by computing numerical fluxes over the
cell interfaces. These, in turn, are obtained from the state
variables, and thus their cell-centered values have to be interpolated
to interface-centered values by appropriate reconstruction
methods. Here, standard schemes are employed, and we refer to the
literature for further details \citep[e.g.,][]{toro2009a}. For the test
problems below, we always used a piecewise linear, MUSCL-like
reconstruction \citep{vanleer1979a, toro2009a} without any limiter because our tests only involve smooth
problems\footnote{At low Mach
    numbers (see Sect.~\ref{sect:lowmach}), well-prepared
    initial conditions are smooth, and this property
    should be retained by the solver. In contrast, 
    discontinuities may occur and require limiting in the transonic or
    supersonic regimes. The implementation of many standard limiters
    is problematic with 
    implicit time discretization approaches. However, for
    flows at high Mach numbers, explicit time discretization is preferred
    in terms of efficiency. Then, standard limiting strategies can be employed.}. The low-Mach-number
methods we focus on here, however, only modify the flux function and can thus be
combined with any kind of reconstruction. In fact, the improvement is
even more significant for low-order schemes because low-Mach
inconsistencies in the flux function are even more pronounced here.

\begin{figure*}
\includegraphics[width=\linewidth]{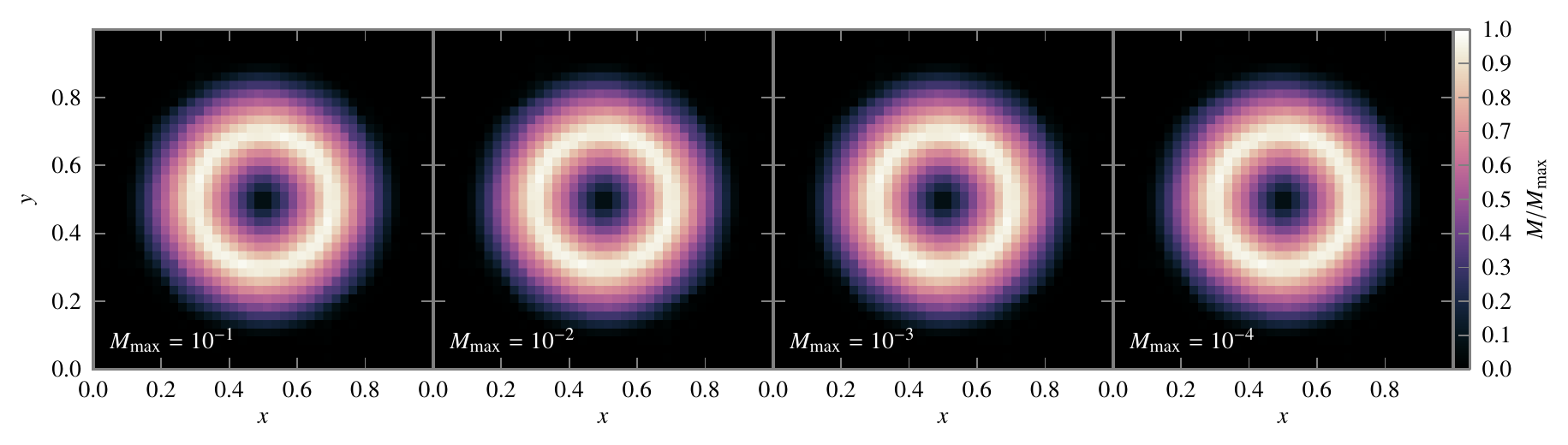}
\caption{Gresho vortex problem advanced to $t=1.0$ with a central flux
  discretization scheme for different maximum Mach
  numbers $M_{\mathrm{max}}$ in the setup, as indicated in the plots. The
  Mach number relative to the respective $M_{\mathrm{max}}$ is
color coded.}
\label{fig:gresho_central}
\end{figure*}

\subsection{Central flux}
Interpolations to interface-centered values, for instance, at the interface
$l+1/2$, with $l\in \lbrace i,j,k \rbrace$, carried out separately
from both ``left'' and ``right'') sides result in different values,
denoted by $U^L_{l+1/2}$ and $U^R_{l+1/2}$\footnote{The
  index $l$ in the following stands for any one of the coordinate
  directions in which we consider the cell interface. The other
  indices are suppressed in this notation}.

A straightforward approach would be to
use \emph{central fluxes},
\begin{equation} \label{eq:central_flux}
\vec{F}_{l+1/2} =  \frac{1}{2}\left( \vec{F}(\vec{U}^L_{l+1/2}) + \vec{F}(\vec{U}^R_{l+1/2})\right).
\end{equation}

As an illustration, we solve the Gresho vortex problem (see
Sect.~\ref{sect:gresho}) with a central flux scheme. The flow at $t=1$
(corresponding to one full revolution of the vortex) is shown in
Fig.~\ref{fig:gresho_central}. Remarkably, it looks almost identical
to the setup at $t=0$ (see Fig.~\ref{fig:gresho_setup}), even for low
values of the maximum Mach number.

However, the scheme resulting from central fluxes is known to be
numerically unstable for many explicit time-discretization schemes. 
In our numerical tests {with implicit time discretization, problems arise
in the evolution of the total kinetic energy of the flow, as
shown in Fig.~\ref{fig:gresho_ekin_central}. The observed increase is
unphysical. Discretization errors continue to grow with time, and therefore a
central flux function cannot be used for practical implementations.
Generally, to achieve stability in the numerical solution of hyperbolic equations, some kind
of upwinding mechanism is necessary. Still, the central-flux
method is instructive for constructing solvers for low Mach numbers, as we
describe below.

\begin{figure}
\includegraphics[width=\linewidth]{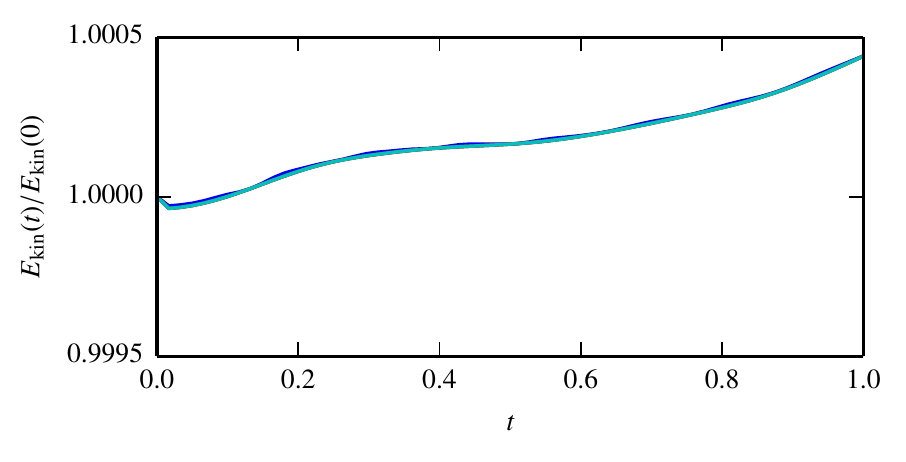}
\caption{Temporal evolution of the total kinetic energy
  $E_{\mathrm{kin}}(t)$ relative to its initial value
  $E_{\mathrm{kin}}(0)$ in the Gresho vortex problem advanced with
  central fluxes. The cases for $M_{\mathrm{max}} = 10^{-1}$,
  $10^{-2}$, $10^{-3}$, and $10^{-4}$ are overplotted but are
virtually
  indistinguishable.}
\label{fig:gresho_ekin_central}
\end{figure}

\subsection{Godunov schemes and Roe flux}
\label{sect:godunov_roe}

\begin{figure*}
\includegraphics[width=\linewidth]{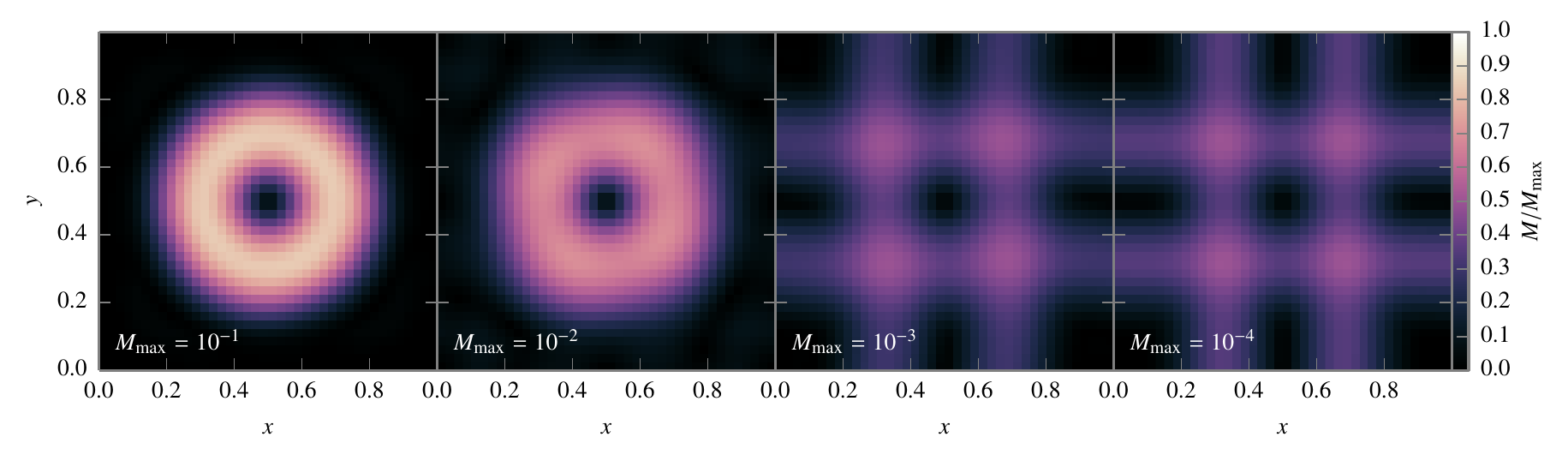}
\caption{Gresho vortex problem advanced to $t=1.0$ with the Roe flux
  discretization scheme for different maximum Mach
  numbers $M_{\mathrm{max}}$ in the setup, as indicated in the plots. The
  Mach number relative to the respective $M_{\mathrm{max}}$ is
color coded.}
\label{fig:gresho_noprecond}
\end{figure*}

One way to achieve stability of numerical schemes is to employ
Godunov-like methods \citep{godunov1959a} for
computing numerical fluxes. Neglecting source terms, the reconstructed
values at cell interfaces pose one-dimensional Riemann problems: At
interface $l+1/2$ we have
\begin{equation}
\label{eq:riemann_1}
  \frac{\partial\fvec{U}}{\partial t} + 
  \frac{\partial\fvec{F}_\xi}{\partial \xi} = 0
\end{equation}
with the initial condition
\begin{equation}
\label{eq:riemann_2}
  \fvec{U}\left( \xi, t=0 \right) \; = \; \left\lbrace
    \begin{array}{l}
      U^L_{l+1/2} \hspace{1em}\mathrm{for}\ \xi>0\\[1ex]
      U^R_{l+1/2} \hspace{1em}\mathrm{for}\ \xi<0.
    \end{array}
  \right.
\end{equation}
The coordinate $\xi$ is measured relative to the cell interface and
along the Cartesian direction $x$, $y$, or $z$ that corresponds to $l
\in \{ i,j,k\}$. The analytic solution to this Riemann
problem is self-similar, 
\begin{equation*}
  \fvec{U}\left( \xi, t \right) = 
  \fvec{U}^{*} \left( \xi / t \right),
\end{equation*}
and can be used to define a numerical flux function,
\begin{equation*}
  \fvec{F}_{l+1/2} = \fvec{F}\left[ \fvec{U}^{*}\left(0\right) \right],
\end{equation*}
over the cell interface $l+1/2$ in $\xi$-direction.

Riemann problems can be solved exactly, but because of inevitable
discretization errors, an approximate solution is usually sufficient
and reduces the computational effort significantly. A very popular
approximate Riemann solver was developed by \citet{roe1981a}. It is based on
a local linearization of Eq.~\eqref{eq:riemann_1},
\begin{equation*}
 \frac{\partial\fvec{U}}{\partial t} + 
  \frac{\partial\fvec{F}_\xi}{\partial \xi} =
  \frac{\partial\fvec{U}}{\partial t} + A (\fvec{U}) \frac{\partial\fvec{U}}{\partial \xi}= 0,
\end{equation*}
 which is achieved by replacing the
flux Jacobian matrix
\begin{equation*}
{A}(\fvec{U}) \equiv \frac{\partial \fvec{F}_\xi}{\partial \fvec{U}} 
\end{equation*}
with a constant matrix $A_\mathrm{roe} (U^L_{l+1/2}, U^R_{l+1/2})$. This
transforms the original Riemann problem
(\ref{eq:riemann_1},\ref{eq:riemann_2}) into a linear system with
constant coefficients while leaving the initial conditions unaltered.
The flux function for this modified system is given by
\citep[e.g.,][]{toro2009a}
\begin{align}
  \label{eq:roeflux-unpre}
  \begin{aligned}
  \fvec{F}_{l+1/2} = \frac{1}{2} & \left[
    \fvec{F}\left( \fvec{U}^L_{l+1/2} \right) +
    \fvec{F}\left( \fvec{U}^R_{l+1/2} \right) \right. \\
&  \left. -
    \left| A_{\mathrm{roe}} \right| \left(
      \fvec{U}^R_{l+1/2} - \fvec{U}^L_{l+1/2}
    \right)
  \right].
\end{aligned}
\end{align}

This expression is used as an approximation to the flux function of
the original nonlinear Riemann problem. 
To this end, the constant matrix $A_\mathrm{roe}$ has to be
specified. Generally, the flux Jacobian matrix $A$ can be decomposed
into a set of right eigenvectors $R$ and a diagonal matrix 
\begin{equation}\label{eq:eigenval}
  \Lambda = \mathrm{diag}\left(
    q_n - \frac{1}{\ndref{M}} c,\ 
    q_n + \frac{1}{\ndref{M}} c,\ 
    q_n, q_n, q_n, q_n
  \right),
\end{equation}
containing the corresponding eigenvalues, so that
\begin{equation}
  A = R \Lambda R^{-1}.
  \label{eq:roe-diag}
\end{equation}
To determine $\left| A_{\mathrm{roe}} \right|$, this matrix and its
absolute value,
\begin{equation}
\label{eq:abs_matrix}
  \left| A \right| \equiv R \left| \Lambda \right| R^{-1},
\end{equation}
are evaluated at the
so-called Roe-averaged state.
\cite{roe1981a} set out three
requirements for its construction. The most important of them,
\begin{equation}
  \label{roe-cons}
  \fvec{F}\left( \fvec{U}^R \right) -
  \fvec{F}\left( \fvec{U}^L \right) =
  A_{\mathrm{roe}} \left(
    \fvec{U}^R - \fvec{U}^L
  \right),
\end{equation}
ensures conservation across discontinuities.

With the three requirements, however, a unique Roe-averaged state can
only be derived for ideal gases. For a general equation of state, such
as needed for astrophysical applications, a number of extensions to
the original Roe scheme have been proposed \citep[see,
e.g.,][]{vinokur1990a,mottura1997a,cinnella2006a}.

In Eq. (\ref{eq:roeflux-unpre}) $|A_{\mathrm{roe}}|$ was introduced from
the solution of a (linearized) Riemann problem at the cell
interface. However, it can be interpreted in a slightly different
fashion. Comparing expression (\ref{eq:roeflux-unpre}) with the
central flux (\ref{eq:central_flux}), we note that the term involving
$|A_{\mathrm{roe}}|$ acts to balance the numerical flux function
either toward the left or right approaching state, depending on the
characteristic decomposition and its eigenvalues. It thus enforces
that the characteristic waves are discretized at the upwind
state of the fluid. This is essential for the numerical stability of
the scheme.

If, for example, the flow is supersonic in the corresponding positive
coordinate direction, all eigenvalues \eqref{eq:eigenval} are
positive. Since, by construction of $A_{\mathrm{roe}}$ relation
(\ref{roe-cons}) holds, the resulting numerical flux is
$\fvec{F}_{l+1/2} = \fvec{F}\left( \fvec{U}^L_{l+1/2} \right)$ and the
other terms in (\ref{eq:roeflux-unpre}) cancel out. For subsonic
flows, the numerical flux is a sum of terms resulting from
discretizing each characteristic wave in the upwind direction.

More generally, the numerical flux function of any Godunov-like scheme
can be expressed in quasi-linear form by
\begin{equation}
\label{eq:godunov_flux}
  \fvec{F}_{l+1/2} = \frac{1}{2} \left[A^L \fvec{U}_{l+1/2}^L + A^R \fvec{U}_{l+1/2}^R -
  D\left(\fvec{U}_{l+1/2}^R - \fvec{U}_{l+1/2}^L\right)\right],
\end{equation}
where $D$ is an ``upwinding matrix''. The corresponding term in the
flux is introduced as a \emph{numerical measure} to stabilize the
scheme by ensuring flux evaluation in upwind direction. When inserted
into the semi-discrete version of the Euler equations
\eqref{eq:finitevolume-disc}, however, it gives rise to the
discretized form of a second space derivative of the state
vector. Thus, it acts as a viscosity, and therefore the corresponding
term is sometimes also called ``upwind artificial viscosity''
\citep{guillard2004a}.  It is clear that it must not dominate the
physical flux represented by the terms involving $A^L$ and $A^R$. This
is the problem encountered for $D = |A_\mathrm{roe}|$ at low Mach
numbers, as we discuss in Sect.~\ref{sect:low-mach-scaling}.

We again tested the performance of Roe's approach to model the
hydrodynamic fluxes by solving the Gresho problem. The result is shown
in Fig.~\ref{fig:gresho_noprecond}. The effect of dissipation is
obvious already for moderately low Mach numbers. After a full
revolution, the Mach number of the flow has decreased and the vortex
does not appear as crisp as in the setup and in the cases with central
fluxes (cf.~Fig.~\ref{fig:gresho_central}). With lower Mach numbers
the dissipation increases strongly, and already at $M_{\mathrm{max}}
\lesssim 10^{-3}$ the vortex is completely dissolved before completing
one revolution.

This behavior can be quantified by the loss of kinetic energy of the
flows, as given in the first row of Table~\ref{tab:ekin-gresho} (``not
preconditioned'').  Although the scheme based on Roe's numerical
fluxes is stable (in contrast to the case of central fluxes where the
kinetic energy increases unphysically), it violates
requirement~\ref{constraint_2} of Sect.~\ref{sect:lowmach}. The
asymptotic behavior of the Euler equations requires the kinetic energy
to be exactly conserved in the zero-Mach number limit. Therefore, the
excessive dissipation of the Roe scheme for low Mach numbers is
attributed to artificial dissipation inherent to the discretization.

\begin{table}
  \caption{\label{tab:ekin-gresho}Percentage of kinetic energy compared
  to the initial value after one revolution ($t=1.0$) of the Gresho vortex for the
  Roe scheme with and without preconditioning. The tests were performed
  using implicit ESDIRK34 time discretization, except for those labeled
  ``(explicit)'', which were made using the time-explicit RK3 scheme.}
  \centering
  \begin{tabular}{r|ccc}
    \hline\hline
    initial max.\ Mach number & $10^{-1}$& $10^{-2}$& $10^{-3}$\\ \hline
    not preconditioned & 94.47 & 74.52 & 49.24\\
    preconditioned & 98.70 & 98.72 & 98.72\\
    preconditioned (explicit) & 97.70 & 97.74 & 98.31\\
    \hline
  \end{tabular}
\end{table}

The failure to reproduce incompressible flow fields is well known in
the fluid dynamics literature \citep[e.g.,][]{volpe1993a,
guillard1999a} and not unique to the particular choice of Roe's
numerical flux function. Instead, \citet{guillard2004a} showed that it
is fundamentally inherent to the Godunov solution strategy. They
examined the scaling behavior of the pressure $p^*$ at cell interfaces
by an expansion of the Riemann problem for the nondimensional Euler
equations in terms of reference Mach number, similar to the approach
discussed in Sect.~\ref{sect:lowmach} [see Eq.~\eqref{eq:exp}]. This
pressure of the self-similar solution to the Riemann problem then
takes the form \citep{guillard2004a}
\begin{equation}
  \label{riemann-pres-asymp}
  p^*=\ndref{\rho} \ndref{c}^2
  \left( p_0 - \frac{1}{2} \ndref{M} \sqrt{\gamma
      p_0 \rho_0 } \Delta u_1 + \cdots \right).
\end{equation}
Obviously, this is in conflict with the scaling behavior in the regime of low
Mach numbers discussed in Sect.~\ref{sect:lowmach}: The pressure
in incompressible flows must be constant in space and time, up to
fluctuations that scale with the \emph{square} of the reference Mach
number. In contrast, $p^*$ contains fluctuations that scale linearly
with $\ndref{M}$, as is characteristic for sound waves. These sound
waves are produced by the first-order velocity jump $\Delta u_1$ over
the interface. Therefore, classical Godunov-like schemes are
inconsistent with requirement~\ref{constraint_1} of
Sect.~\ref{sect:lowmach}.
 
The fact that an incompressible solution cannot be maintained by
standard Godunov-like methods does not come as a surprise. We
recall that
jumps in the flow variables are intentionally created at the cell
interfaces by the Godunov discretization strategy in order to define
Riemann problems that can easily be solved. This is in conflict with
the smooth flow-fields in the incompressible regime. The artificial
introduction of discontinuities leads to the generation of
\emph{artificial sound waves,} which are not resolved on the grid and
quickly get dissipated. This provides a heuristic argument for the
excessive dissipation of the schemes at low Mach numbers.  It may
therefore be argued that the Godunov approach is unsuited for
modeling flows at low Mach numbers. However, for reasons given in
Sect.~\ref{sect:lowmach}, such schemes are still attractive for
a numerical simulation of astrophysical problems. As a measure to
salvage conventional solvers, preconditioning techniques for the
numerical flux functions have been proposed \citep[see][for a
review]{turkel1999a} , and we investigate this approach in the
following.

\section{Low-Mach-number scaling of Godunov-type numerical flux
  functions}
\label{sect:low-mach-scaling}

The differences in the low-Mach-number behavior observed in the
central flux and the Roe schemes indicate the origin of the
excessive dissipation in the latter. Comparing the flux functions of
the central scheme (\ref{eq:central_flux}) to that of Roe
(\ref{eq:roeflux-unpre}), or, more generally, to the flux function of
any Godunov-like scheme (\ref{eq:godunov_flux}), we suspect the
respective upwinding terms
\begin{equation*}
\left| A_{\mathrm{roe}} \right| \left(
      \fvec{U}^R_{l+1/2} - \fvec{U}^L_{l+1/2}
    \right)
\end{equation*}
or
\begin{equation*}
D\left(\fvec{U}_{l+1/2}^R - \fvec{U}_{l+1/2}^L\right)
\end{equation*}
to be the origin of the numerical dissipation.

It is therefore essential to analyze the scaling with respect to the
reference Mach number $\ndref{M}$ of the matrix elements $A(\fvec{U})$
and $D$ in the numerical flux function (\ref{eq:godunov_flux}). Since
for the Euler equations $\fvec{F}_{\xi} = A(\fvec{U}) \fvec{U}$ and D
occurs in the numerical flux of Godunov-like schemes, it is important
that their scalings with reference Mach number be equal.

We present a scaling analysis for the Roe flux, inspired by the
discussion of \cite{turkel1999a}.  The matrices $A$ and $D$ take a
particularly simple and convenient form when expressed in terms of
primitive variables,
\begin{equation}
  \label{eq:def-V}
  \fvec{V} = (\rho, u, v, w, p, X)^T.
\end{equation}
The scaling of $A$ can easily be determined by taking the derivative
of the flux vector, as defined in Eq.~(\ref{eq:flux-general}), with
respect to $\fvec{V}$ and multiplying it from the left with the
transformation matrix $\frac{\partial\fvec{V}}{\partial\fvec{U}}$,
\begin{align}
\begin{aligned}
  A_\fvec{V} =& \frac{\partial \fvec{V}}{\partial \fvec{U}}
  \frac{\partial \fvec{F}}{\partial \fvec{U}} \frac{\partial \fvec{U}}{\partial
    \fvec{V}}=\\ & 
    \begin{pmatrix}
      q_n & \rho n_x     & \rho n_y     & \rho n_z     & 0                                  & 0\\[1ex]
      0   & q_n          & 0            & 0            & \frac{n_x}{\rho \ndref{M}^2} & 0\\[1ex]
      0   & 0            & q_n          & 0            & \frac{n_y}{\rho \ndref{M}^2} & 0\\[1ex]
      0   & 0            & 0            & q_n          & \frac{n_z}{\rho \ndref{M}^2} & 0\\[1ex]
      0   & \rho c^2 n_x & \rho c^2 n_y & \rho c^2 n_z & q_n                                & 0\\[1ex]
      0   & 0            & 0            & 0            & 0                                  & q_n
    \end{pmatrix}.
\end{aligned}
\label{eq:av}
\end{align}
The expressions for the transformation matrices are given in
Appendix~\ref{sect:var-trans}.  The scaling with the reference Mach
number can be read off directly from (\ref{eq:av}) because all
nondimensional variables are of order one:
\begin{equation}
  \label{eq:scaling-fluxjac}
  A_{\fvec{V}} \propto \left(
    \begin{array}{cccccc}
      \ord{1} & \ord{1} & \ord{1} & \ord{1} & 0 & 0 \\[1ex]
      0 & \ord{1} & 0 & 0 & \ord{\frac{1}{\ndref{M}^2}} & 0 \\[1ex]
      0 & 0 & \ord{1} & 0 & \ord{\frac{1}{\ndref{M}^2}} & 0 \\[1ex]
      0 & 0 & 0 & \ord{1} & \ord{\frac{1}{\ndref{M}^2}} & 0 \\[1ex]
      0 & \ord{1} & \ord{1} & \ord{1} & \ord{1} & 0 \\[1ex]
      0 & 0 & 0 & 0 & 0 & \ord{1}
    \end{array}
  \right).
\end{equation}

Next, we apply the same kind of analysis to the upwind matrix of the
Roe solver, $\left| A_{\mathrm{roe}} \right|$ in
Eq.~(\ref{eq:roeflux-unpre}). The following calculations were
performed using computer algebra software. We outline the procedure
and give intermediate results.

Starting out from the flux Jacobian in primitive variables,
Eq.~(\ref{eq:av}), we compute its eigenvalues,
\begin{equation*}
  \Lambda = \text{diag}(q_n - \frac{c}{\ndref{M}}, q_n + \frac{c}{\ndref{M}}, q_n, q_n, q_n, q_n),
\end{equation*}
and corresponding left and right eigenvectors, $R$ and $R^{-1}$. In
the Roe scheme, the absolute value of all components of $\Lambda$ is
taken [see Eq.~\eqref{eq:abs_matrix}].  To analyze the behavior of the
Roe matrix, we distinguish different cases depending on the absolute
value of the fluid velocity. The supersonic case ($q_n> c/\ndref{M}$ or
$q_n<-c/\ndref{M}$) is simple because the eigenvalues are either all
positive or all negative. Thus, 
\begin{equation*} 
\left| A \right| = \begin{cases} R \left| \Lambda \right| R^{-1} =  R
  \Lambda R^{-1} = A, & \mbox{if } q_n > \frac{c}{\ndref{M}},\\
 R \left| \Lambda \right| R^{-1} =  R (-\Lambda) R^{-1} = -A, &
 \mbox{if } q_n<
-\frac{c}{\ndref{M}}.
\end{cases}
\end{equation*}
The scaling of $|A|$ with $\ndref{M}$ is
the same as in the analytical case, Eq.~\eqref{eq:scaling-fluxjac}, as
the matrix is just multiplied by a scalar.

The subsonic case ($c/\ndref{M} > q_n > -c/\ndref{M}$)
is more involved. Without loss of generality, we assume $0 < q_n <
c/\ndref{M}$, which means that only the sign of the eigenvalue $q_n -
\frac{c}{\ndref{M}}$ has to be changed. Multiplying with $R$ and
$R^{-1}$ yields the matrix
\begin{equation*}
\left| A_{\fvec{V}} \right| \equiv  R \left| \Lambda \right| R^{-1}=
  \begin{pmatrix}
    q_n & \alpha n_x &\alpha n_y  &\alpha n_z & \frac{\frac{c}{\ndref{M}} - q_n}{c^2}\\[1ex]
    0   & \phi_x  & n_x n_y \beta & n_x n_z \beta & \frac{n_x q_n}{c \rho \ndref{M}} & 0\\[1ex]
    0   & n_x n_y \beta & \phi_y & n_y n_z \beta & \frac{n_y q_n}{c \rho \ndref{M}} & 0\\[1ex]
    0   & n_x n_z \beta & n_y n_z \beta & \phi_z & \frac{n_z q_n}{c \rho \ndref{M}} & 0\\[1ex]
    0   & c^2 \alpha n_x & c^2 \alpha n_y & c^2 \alpha n_z & \frac{c}{\ndref{M}} & 0\\[1ex]
    0 & 0 & 0 & 0 & 0 & q_n
  \end{pmatrix},
\end{equation*}
with $\alpha = \rho \ndref{M} q_n / c$, $\beta = (c - \ndref{M} q_n) /
\ndref{M}$, and $\phi_i = \frac{c n_i^2}{\ndref{M}} + (1 - n_i^2)
q_n$.
The scaling of $\left| A_{\fvec{V}} \right|$ with the Mach number is then given by
\begin{equation*}
\left| A_{\fvec{V}} \right| \propto
 \begin{pmatrix}
    \ord{1} & \ord{\ndref{M}}           & \ord{\ndref{M}}           & \ord{\ndref{M}}           & \ord{\frac{1}{\ndref{M}}} & 0 \\[1ex]
    0 & \ord{\frac{1}{\ndref{M}}} & \ord{\frac{1}{\ndref{M}}} & \ord{\frac{1}{\ndref{M}}} & \ord{\frac{1}{\ndref{M}}} & 0 \\[1ex]
    0 & \ord{\frac{1}{\ndref{M}}} & \ord{\frac{1}{\ndref{M}}} & \ord{\frac{1}{\ndref{M}}} & \ord{\frac{1}{\ndref{M}}} & 0 \\[1ex]
    0 & \ord{\frac{1}{\ndref{M}}} & \ord{\frac{1}{\ndref{M}}} & \ord{\frac{1}{\ndref{M}}} & \ord{\frac{1}{\ndref{M}}} & 0 \\[1ex]
    0 & \ord{\ndref{M}}           & \ord{\ndref{M}}           & \ord{\ndref{M}}           & \ord{\frac{1}{\ndref{M}}} & 0 \\[1ex]
    0 & 0                   & 0                   & 0                   & 0                   & \ord{1}
  \end{pmatrix}.
\end{equation*}
This result is clearly inconsistent with the scaling of the physical
flux Jacobian in Eq.~\eqref{eq:scaling-fluxjac}. 

This behavior is undesired because the only purpose of the upwind term
involving $|A|$ is to provide a balance between the flux function
evaluated at both sides of the interface. Instead of fulfilling this
purpose, many elements of the upwinding matrix dominate the
physical flux at low Mach numbers as a result of the incorrect scaling
behavior. In particular, the $\ord{1/\ndref{M}}$ terms in the velocity
equations (corresponding to rows 2 to 4 of $\left| A_{\fvec{V}}
\right|$) can be identified as an origin of the observed problems.
Therefore, the dominant dissipative character of the upwind artificial
viscosity term prevents the correct representation of the physical
flows at low Mach numbers\footnote{Note, however, that such problems
  are not restricted to upwind artificial viscosity terms that occur
  in many approximate Riemann solvers. \cite{guillard2004a} showed that
  they generally occur in Godunov schemes. }.

\section{Preconditioned numerical fluxes for the regime of low Mach numbers}
\label{sect:preconditioner}

To cure the inconsistent scaling behavior of numerical flux functions,
the upwind artificial viscosity terms are modified by multiplication
with preconditioning matrices.  Originally, such matrices were
used to accelerate convergence of numerical schemes to
steady-state solutions of the Euler equations. Here,
$\partial\fvec{U}/\partial t = 0$. To find such solutions numerically,
however, the time dependence is retained (or replaced with some
fictitious pseudo-time dependence) and used to march the system to a
steady state in an iterative process. This is complicated by the
stiffness of the system at low Mach numbers, which can be reduced by
multiplying the time derivative of the state vector with a suitable
invertible matrix,
\begin{equation*}
  P^{-1} \frac{\partial\fvec{U}}{\partial t}.
\end{equation*}
This modification leaves the steady-state solution unchanged as the
time derivative vanishes here and is only used for the numerical
solution process. In one dimension, this approach transforms
system \eqref{eq:euler} into 
\begin{alignat*}{5}
  P^{-1}& \frac{\partial \fvec{U}}{\partial t} \,+\,&&\frac{\partial
  \fvec{F}_x}{\partial x} &= 0,\\
  &\frac{\partial \fvec{U}}{\partial t} \, +  &P&\frac{\partial
  \fvec{F}_x}{\partial x} &= 0,
\end{alignat*}
or, in quasi-linear form,
\begin{equation}
\label{eq:precond_steady}
 \frac{\partial \fvec{U}}{\partial t} + P A(\fvec{U}) \frac{\partial \fvec{U}}{\partial x}
 = 0,
\end{equation}
with $A(\fvec{U}) = {\partial \fvec{F}_x}/{\partial \fvec{U}}$.

The matrix $P$ is constructed to reduce the stiffness of the system at
low Mach numbers by equalizing the eigenvalues of the modified flux
Jacobian $P A(\fvec{U})$. This reduces the condition number
\begin{equation*}
\kappa = \max{\frac{|\lambda_i|}{|\lambda_j|}}
\end{equation*}
over all eigenvalues $\lambda$ of the matrix, which the convergence
rate depends on, and hence is called ``preconditioning''
\citep[see][for an overview of this technique]{turkel1999a}.

Similar to what was discussed in Sect.~\ref{sect:num_flux}, an
artificial viscosity has to be introduced in Eq. \eqref{eq:precond_steady}
to stabilize numerical solution schemes. For a Roe solver, for
example, this results in the system
\begin{equation}
\label{eq:precond_steady_visc}
\frac{\partial \fvec{U}}{\partial t} + P _\mathrm{roe} A(\fvec{U}) \frac{\partial
  \fvec{U}}{\partial x} = \frac{1}{2} \frac{\partial}{\partial x}\left( \left| PA
  \right|_\mathrm{roe} \frac{\partial \fvec{U}}{\partial x}\right),
\end{equation}
where the preconditioning matrix is evaluated at the Roe state.

Apart from accelerating convergence to steady state, it was observed
that preconditioning also improves the accuracy of the solutions
\citep[e.g.,][]{turkel1993a, turkel1999a}.  Multiplication of
Eq.~\eqref{eq:precond_steady_visc} from the left with
$P^{-1}_\mathrm{roe}$,
\begin{equation*}
P^{-1}_\mathrm{roe}\frac{\partial \fvec{U}}{\partial t} + A(\fvec{U}) \frac{\partial
  \fvec{U}}{\partial x} = \frac{1}{2} \frac{\partial}{\partial x}\left[ \left(P^{-1}\left| PA
  \right|\right)_\mathrm{roe} \frac{\partial \fvec{U}}{\partial x}\right],
\end{equation*}
shows that both improvements are due to modifications of the two terms
in the system that were introduced for numerical reasons, that
is, the
time derivative of the state vector and the viscosity in
Eq.~\eqref{eq:precond_steady_visc}. The spatial derivative of the flux
remains unchanged. \citet{turkel1999a} pointed out that in principle,
the preconditioning matrices in the two terms can be different and
chosen to serve their individual purposes best:
\begin{equation*}
Q^{-1}_\mathrm{roe}\frac{\partial \fvec{U}}{\partial t} + A(\fvec{U}) \frac{\partial
  \fvec{U}}{\partial x} = \frac{1}{2} \frac{\partial}{\partial x}\left[ \left(P^{-1}\left| PA
  \right|\right)_\mathrm{roe} \frac{\partial \fvec{U}}{\partial x}\right],
\end{equation*}
with some preconditioning matrices $P$ and $Q$.  This is possible
since these terms are artificial, that is, they are not part of the original
system to be solved.

In contrast, for time-dependent flows}, a
modification of the time derivative has to be avoided because it now
has physical significance. Therefore $Q$ is set to the identity
matrix. To improve the solution in the regime of low Mach numbers,
however, preconditioning with $P$ is still applied to the upwind
artificial viscosity term of the Roe scheme. Thus, for time-dependent
problems, the original equations are discretized as in
Eq.~\eqref{eq:finitevolume-disc}.  The only difference to the original
scheme is that the numerical
fluxes are calculated from the Riemann problem involving the preconditioned system
\begin{equation*}
  \frac{\partial\fvec{U}}{\partial t} + 
  P \frac{\partial\fvec{F}_\xi}{\partial \xi} = 0
\end{equation*}
instead of Eq. \eqref{eq:riemann_1}, without changing the initial
conditions given in Eq.~\eqref{eq:riemann_2}. This results in the
preconditioned numerical flux function
\begin{multline}
  \label{eq:roeflux-pre}
  \fvec{F}_{l+1/2} = \frac{1}{2} \left[
  \fvec{F}\left( \fvec{U}^L_{l+1/2} \right) +
  \fvec{F}\left( \fvec{U}^R_{l+1/2} \right) -\right.\\
  \left. \left(P^{-1}\left| \left\{P A\right\}_{\mathrm{roe}} \right| \right)\left(
  \fvec{U}^R_{l+1/2} - \fvec{U}^L_{l+1/2}
  \right)
  \right].
\end{multline}

We again emphasize that while the upwind artificial viscosity term is
modified, the discretization of the temporal terms and the central
flux remains unchanged. Therefore, only the \emph{numerical} dissipation is
altered with respect to the unpreconditioned solver. If an appropriate
preconditioning matrix is specified, the same physical flow can be
described more accurately by the numerical solution.

\begin{figure*}
\includegraphics[width=\linewidth]{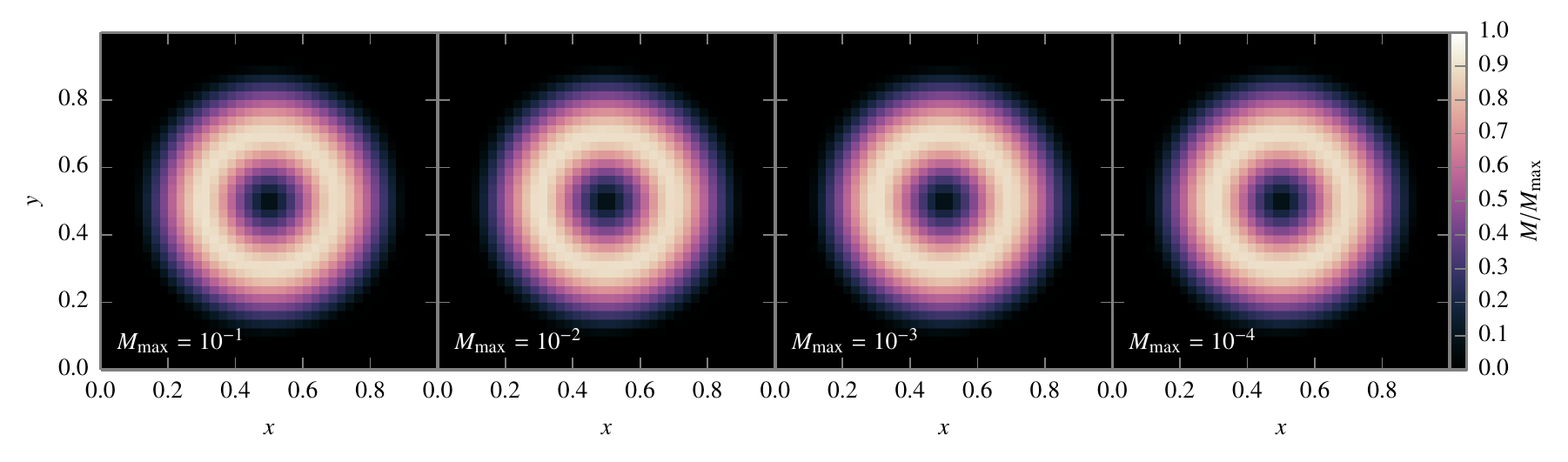}
\caption{Gresho vortex problem advanced to $t=1.0$ with preconditioned Roe
  fluxes for different maximum Mach
  numbers $M_{\mathrm{max}}$ in the setup, as indicated in the plots. The
  Mach number relative to the respective $M_{\mathrm{max}}$ is
color coded.}
\label{fig:gresho_precond}
\end{figure*}

A popular choice for a preconditioning matrix was introduced
by \citet{weiss1995a}.  Expressed in primitive variables, it has the form
\begin{equation}
\label{eq:pc_weiss}
  P_\vec{V} = 
  \begin{pmatrix}
    1 & 0 & 0 & 0 & \frac{\mu^2 - 1}{c^2} & 0\\[1ex]
    0 & 1 & 0 & 0 & 0 & 0\\[1ex]
    0 & 0 & 1 & 0 & 0 & 0\\[1ex]
    0 & 0 & 0 & 1 & 0 & 0\\[1ex]
    0 & 0 & 0 & 0 & \mu^2 & 0\\[1ex]
    0 & 0 & 0 & 0 & 0 & 1
  \end{pmatrix},
\end{equation}
with the parameter $\mu = \min[1,\max(M,M_\text{cut})]$, which should
be set to the \emph{local} Mach number $M$ of the fluid (e.g.,
numerically defined the average of both sides of the interface). Its lower limit, $M_\text{cut}$, avoids
singularity of $P_\vec{V}$. To
prevent the preconditioner from acting on supersonic flow, $\mu$ is
limited from above to $1$. At locations in the flow where $M \ge 1 $,
the preconditioner reduces to the identity matrix, and the original Roe
scheme is recovered.

For the preconditioner \eqref{eq:pc_weiss}
we performed the same low-Mach analysis as in
Sect.~\ref{sect:low-mach-scaling}. This yields
\begin{equation*}
  P^{-1}_{\fvec{V}} \left| P_{\fvec{V}} A_{\fvec{V}} \right| 
  \propto
  \begin{pmatrix}
    \ord{1} & \ord{1} & \ord{1} & \ord{1} & \ord{\frac{1}{\ndref{M}^2}} & 0 \\[1ex]
    0       & \ord{1} & \ord{1} & \ord{1} & \ord{\frac{1}{\ndref{M}^2}} & 0 \\[1ex]
    0       & \ord{1} & \ord{1} & \ord{1} & \ord{\frac{1}{\ndref{M}^2}} & 0 \\[1ex]
    0       & \ord{1} & \ord{1} & \ord{1} & \ord{\frac{1}{\ndref{M}^2}} & 0 \\[1ex]
    0       & \ord{1} & \ord{1} & \ord{1} & \ord{\frac{1}{\ndref{M}^2}} & 0 \\[1ex]
    0       & 0       & 0       & 0       & 0                           & \ord{1}
  \end{pmatrix}.
\end{equation*}
Obviously, the scaling behavior of the preconditioned Roe matrix is
almost perfectly consistent with the analytical scaling in
Eq.~\eqref{eq:scaling-fluxjac}. The only discrepancy is the fifth
column in rows one and five, corresponding to an increased mass and
energy flux in the low-Mach regime depending on the local change in
pressure. This is particularly problematic in setups of stellar
astrophysics, which almost always involve a hydrostatic stratification
with a pressure gradient in the vertical direction. Terms in the flux
function resulting from elements of the upwind matrix with incorrect
scaling behavior make it then impossible to maintain the correct
hydrostatic equilibrium.

We thus propose a new low-Mach preconditioning
scheme for the Roe flux that ensures the correct scaling behavior in
slow flows. This is achieved by using the preconditioning
matrix
\begin{equation}
  \label{eq:flux-precond-matrix}
  P_\vec{V} = 
  \begin{pmatrix}
      1 & 
      n_x \frac{\rho \delta \ndref{M}}{c} &
      n_y \frac{\rho \delta \ndref{M}}{c} &
      n_z \frac{\rho \delta \ndref{M}}{c} & 0 & 0 \\[1ex]
      0 & 1 & 0 & 0 & -n_x \frac{\delta}{\rho c \ndref{M}} & 0 \\[1ex]
      0 & 0 & 1 & 0 & -n_y \frac{\delta}{\rho c \ndref{M}} & 0 \\[1ex]
      0 & 0 & 0 & 1 & -n_z \frac{\delta}{\rho c \ndref{M}} & 0 \\[1ex]
      0 &
      n_x \rho c \delta \ndref{M} &
      n_y \rho c \delta \ndref{M} &
      n_z \rho c \delta \ndref{M} & 1 & 0 \\[1ex]
      0 & 0 & 0 & 0 & 0 & 1
  \end{pmatrix},
\end{equation}
with the parameter $\delta = 1/\mu - 1$. As in
Eq.~\eqref{eq:pc_weiss}, $\mu$ is the local Mach number limited by 1
and $M_\text{cut}$. The unpreconditioned case corresponds to
$\delta=0$, which is automatically reached in regions with local Mach
numbers of $1$ or larger. Here, the scheme transitions continuously to
the original Roe scheme.

With this scheme, an analysis of the low-Mach-number scaling behavior
gives
\begin{equation*}
  P^{-1}_{\fvec{V}} \left| P_{\fvec{V}} A_{\fvec{V}} \right| 
  \propto
    \begin{pmatrix}
      \ord{1} & \ord{1} & \ord{1} & \ord{1} & 
      \ord{1} & 0 \\[1ex]
      0 & \ord{1} & \ord{1} & \ord{1} & 
      \ord{\frac{1}{\ndref{M}^2}} & 0 \\[1ex]
      0 & \ord{1} & \ord{1} & \ord{1} & 
      \ord{\frac{1}{\ndref{M}^2}} & 0 \\[1ex]
      0 & \ord{1} & \ord{1} & \ord{1} & 
      \ord{\frac{1}{\ndref{M}^2}} & 0 \\[1ex]
      0 & \ord{1} & \ord{1} & \ord{1} & 
      \ord{1} & 0 \\[1ex]
      0 & 0 & 0 & 0 & 0 & \ord{1}
    \end{pmatrix}.
\end{equation*}
This shows that the new scheme behaves consistently with the physical
flux Jacobian matrix. None of the terms is overwhelmed by the numerical
upwind artificial viscosity for $\ndref{M} \rightarrow 0$, as was the case for the
unpreconditioned Roe scheme and the scheme preconditioned with matrix
\eqref{eq:pc_weiss}.  

We remark, however, that our new flux function is not unique -- there
may be other ways of constructing preconditioning matrices to correct
the low-Mach-number scaling of the Roe flux. It is also not proved
that our scheme is numerically stable.  On the basis of the Gresho
vortex problem and the Kelvin-Helmholtz instability (see
Sect.~\ref{sect:tests}), however, we demonstrate that the scheme has
the desired behavior.

\section{Tests of the low-Mach-number solver}
\label{sect:tests}

\subsection{Gresho vortex problem}

We evolved the Gresho vortex problem from $t=0$ to $t=1$ with our new
low-Mach-number solver, preconditioned with matrix
\eqref{eq:flux-precond-matrix}.  Results for maximum Mach numbers of
$M_{\mathrm{max}} = 10^{-n}$, $n = 1, 2, 3, 4$ are shown in
Fig.~\ref{fig:gresho_precond}, but tests down to $n= 10$ did not
detect any deviation from the illustrated behavior. As before, all
these runs were performed with implicit ESDIRK34 time discretization
using the advective CFL criterion \eqref{eq:cfl_adv}. Because the
modifications introduced in our new method are in the spatial
discretization alone, we expect it to work for explicit time
discretization as well as for implicit schemes. To demonstrate this,
we performed some of the tests using the explicit RK3 scheme
\citep{shu1988a}. While for low Mach numbers this is computationally
very inefficient compared to the implicit case, the amount of
dissipation is almost identical (see Table~\ref{tab:ekin-gresho}). In
our tests we found no time-step restriction other than the usual
acoustic CFL condition that is derived from the physical signal
velocities.

Our tests tests impressively demonstrate the performance of the new
solver. The representation of the Gresho vortex does not degrade with
lower Mach numbers, even if it is decreased by nine orders of magnitude.
Visually, no differences can be found in the plots shown in
Fig.~\ref{fig:gresho_precond}, and this trend continues to even lower
Mach numbers.

\begin{figure}[h]
  \includegraphics[width=\linewidth]{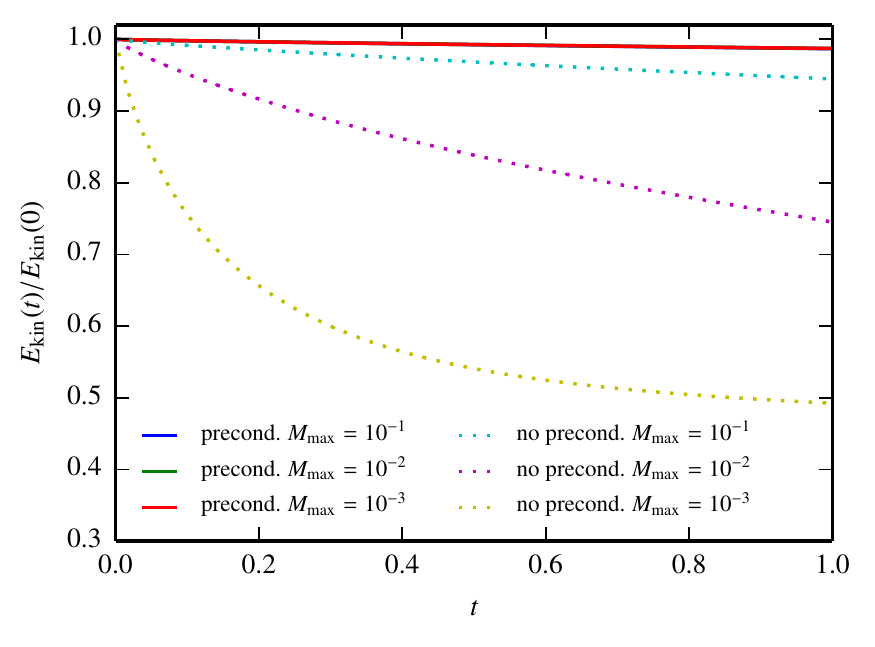}
  \caption{\label{fig:gresho_ekin_compare}Temporal evolution of the total kinetic energy
  $E_{\mathrm{kin}}(t)$ relative to its initial value
  $E_{\mathrm{kin}}(0)$ in the Gresho vortex problem. For different
  initial Mach numbers simulations were run with (solid) and without
  (dotted) flux preconditioning. The lines for the preconditioned scheme
  show very low dissipation and are virtually indistinguishable, that is,
  independent of the Mach number. The final values are listed in
  Table~\ref{tab:ekin-gresho}.}
\end{figure}

This is confirmed in Fig.~\ref{fig:gresho_ekin_compare}, where the
evolution of the total kinetic energy is shown over one full
revolution of the vortex for different setups and compared to the
results obtained without preconditioning. No signs of instability of
our scheme are detected, and only $\le 2.3$\% of the initial energy is
lost over the simulation time for all cases with
preconditioning (see the corresponding rows of Table~\ref{tab:ekin-gresho},
labeled ``preconditioned''). In contrast, already at a moderately
low initial Mach number of $10^{-1}$, the unpreconditioned solver
loses more than $5.5$\% of the kinetic energy in the same setup. This
shows that the numerical dissipation of the scheme is generally reduced
by preconditioning.

At lower Mach numbers, the loss of kinetic energy drastically increases for the
unpreconditioned method, while it is virtually unchanged for the
preconditioned case. Thus, the second (and most important) observation
is that the evolution of the kinetic energy is independent of the
maximum Mach numbers $M_{\mathrm{max}}$ in the setup when applying our
low-Mach-number preconditioner. We thus conclude that our new solver
matches the design goals of a Mach-number-independent dissipation.

\subsection{Kelvin--Helmholtz instability}
\label{sect:kh2d}

\begin{figure*}
  \centering
  \includegraphics[width=\textwidth]{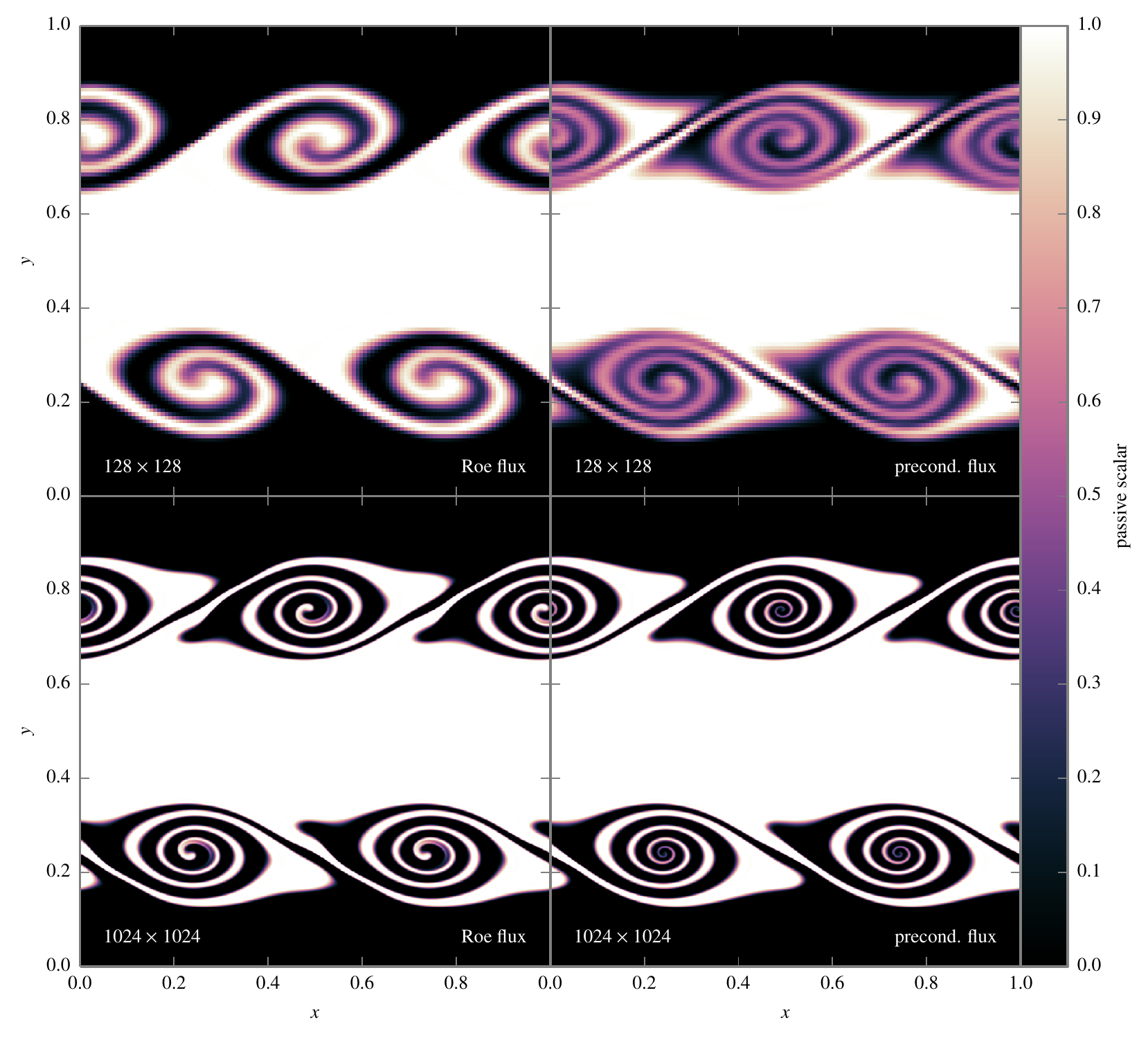}
  \caption{\label{fig:kh2d}Advection of a passive scalar in the setup
  described in Sect.~\ref{sect:kh2d}. The left panels were computed
  using the unmodified Roe scheme. In the right panels, the flux
  preconditioning method from Eq.~\eqref{eq:roeflux-pre} was used with the
  matrix from Eq.~\eqref{eq:flux-precond-matrix}. The top panels show
  low-resolution simulations ($128^2$), the bottom panels show the same
  setup at higher resolution ($1024^2$). All panels show the
  state at a nondimensional time of $3.0$. Initially, the central (rightward-flowing) region
  had a value of $X = 1$ and the upper and lower (leftward-flowing) regions had a
  value of $X = 0$.}
\end{figure*}

Although the Gresho vortex provides a useful testbed for the
performance of numerical hydrodynamics solvers at
low Mach numbers, its solution is essentially a steady
state because of its symmetry. A true time-dependent test problem that is also closer to
astrophysical application is the Kelvin--Helmholtz instability,
which arises at the interface of two shearing flows. Its analytic theory was
set out by \citet{chandrasekhar1961a}. There,
however, the initial conditions were arranged using a sharp
discontinuity. This is unsuitable for comparing different numerical
discretizations and resolutions because the time and place of the
initial development of instabilities entirely depend on random
numerical effects. To prevent this, we instead employed the setup of
\citet{mcnally2012a}, which provides smooth initial conditions and an
initial perturbation perpendicular to the interface between the
shearing flows.

To solve the homogeneous Euler equations, we chose a computational
domain of $[0,1]\times[0,1]$ with periodic boundaries in both
directions. The problem was set up in nondimensional variables so that
the Mach number could easily be adjusted. The domain was split into four
regions, the leftward-moving upper ($y>0.75$) and lower ($y<0.25)$ parts,
and the rightward-moving center ($0.25<y<0.75$), which, in turn, was split
into two regions to enable a smooth transition. The initial conditions
were
\begin{align*}
  p &= 2.5,\\[1ex]
  \rho &=
  \begin{cases}
    \rho_1 - \rho_\text{m} e^{\frac{y-0.25}{L}},    & \mbox{if } 0.25 > y \geq 0,\\
    \rho_2 + \rho_\text{m} e^{\frac{-y+0.25}{L}},   & \mbox{if } 0.5 > y \geq 0.25,\\
    \rho_2 + \rho_\text{m} e^{\frac{-(0.75-y)}{L}}, & \mbox{if } 0.75 > y \geq 0.5,\\
    \rho_1 - \rho_\text{m} e^{\frac{-(y-0.75)}{L}}, & \mbox{if } 1 > y \geq 0.75,
  \end{cases}\\[1ex]
  u &=
  \begin{cases}
    u_1 - u_\text{m} e^{\frac{y-0.25}{L}},    & \mbox{if } 0.25 > y \geq 0,\\
    u_2 + u_\text{m} e^{\frac{-y+0.25}{L}},   & \mbox{if } 0.5 > y \geq 0.25,\\
    u_2 + u_\text{m} e^{\frac{-(0.75-y)}{L}}, & \mbox{if } 0.75 > y \geq 0.5,\\
    u_1 - u_\text{m} e^{\frac{-(y-0.75)}{L}}, & \mbox{if } 1 > y \geq 0.75.
  \end{cases}
\end{align*}
We fixed the parameters to $\rho_1 = 1.0$, $\rho_2 = 2.0$, $\rho_\text{m} = (\rho_1 -
\rho_2) / 2$, $u_1=0.5$, $u_2=-0.5$, $u_\text{m} = (u_1 - u_2) / 2$, and
$L=0.025$. The equation of state is that of an ideal gas with
$\gamma=5/3$.

The instability is seeded with an initial perturbation in the transverse
velocity given by
\begin{equation*}
  v = 10^{-2} \sin(2 \pi x).
\end{equation*}
This results in a single mode being excited.

To adjust the Mach number of this setup to our needs, we set the
reference quantities to $\ndref{\rho}=1$, $\ndref{p}=2.5$,
$\ndref{x}=1$. The reference Mach number~$\ndref{M}$ is used as a
parameter to prescribe the Mach number of the flow. The other reference
quantities are then computed from this minimal set of reference
quantities using $\ndref{c} = \sqrt{\ndref{p}/\ndref{\rho}}$ and
$\ndref{u} = \ndref{M} \ndref{c}$.

With this initial setup, we carried out simulations at Mach number
$10^{-2}$, one with the original (unpreconditioned) Roe scheme and one with flux
preconditioning according to our matrix \eqref{eq:flux-precond-matrix}. For each case we performed a
low-resolution run for which the domain was discretized on a
numerical grid of $128^2$ cells, and a high-resolution run with
$1024^2$ cells. A passive scalar
quantity $X$ was advected with the fluid flow for better illustration of the
mixing of the layers. Its initial value is $X=1$ in the center ($0.25\leq y
< 0.75$) and $X = 0$
outside. All runs used second-order reconstruction and the implicit
ESDIRK34 time-stepping scheme with an advective CFL criterion according
to Eq.~\eqref{eq:cfl_adv}.

The results at nondimensional time $3.0$ are shown
in Fig.~\ref{fig:kh2d}. As we solve the homogeneous Euler equations
without any diffusion, all mixing that takes place is solely due to
numerical errors. Ideally, every turnover of the instability should
remain visible in the simulation because the interface is distorted, but no
mixing occurs. This is clearly the case for the high-resolution runs
shown in the bottom row of Fig.~\ref{fig:kh2d}.

The high-resolution runs are very similar and differ only in the
innermost part of the vortex. Here, the simulation with the
unpreconditioned Roe flux shows some diffusive effects, which are
absent for the preconditioned flux. In case of the
low-resolution runs the difference is much more pronounced. With flux
preconditioning the solution is qualitatively similar to the
high-resolution runs. The unmodified Roe flux significantly smears out the flow
and results in a
different morphology than the other runs. This example illustrates that
while the problems of standard flux functions at low Mach numbers can be partially
alleviated by higher resolution, they still appear close to the grid
scale. From a different perspective, one can state that use of
the preconditioned flux at low resolution helps to
reproduce results that would normally require much better resolution and
thus more computational resources.

\subsection{Gresho vortex combined with a sound wave}

The previous two test problems are both tests that reside strictly in
the low-Mach, nearly incompressible regime. While the preconditioned
Roe scheme handles these cases well, its unique strength lies in
handling of situations were compressible and incompressible conditions
coexist on the same grid. To test the performance, we constructed a new
problem using the Gresho vortex as a basis. The domain was extended to
the left and comprised $80\times40$ grid cells. In the $40\times40$
cells on the right-hand side of the domain, the Gresho vortex was set
up as described in Sect.~\ref{sect:gresho} with $M_\mathrm{max} =
10^{-3}$.  In the $40\times40$ cells to the left of it,
we placed a plane sound wave packet propagating to the right toward
the vortex. The initial condition of the sound wave packet is given by
\begin{align*}
  \psi &= \exp\left(-\alpha(x-x_0)^2\right)/\ndref{M},\\
  u_1 &= - 2 \alpha (x-x_0) \psi,\\
  p_1 &= -2 \alpha c \ndref{M} (x-x_0) \psi,\\
  \rho_1 &= p_1 / c^2,
\end{align*}
with $\alpha = (0.1)^{-2}$ controlling the width of the packet,
$x_0=-0.5$ denoting the position of the center, and $c = \sqrt{\gamma p_0 /
\rho_0}$ the sound speed. The nondimensional values $u_1$, $p_1$, and
$\rho_1$ were added to the background states of the Gresho vortex
problem. This resulted in a sound wave with a maximum Mach number of
$10^{-2}$.

\begin{figure*}
\includegraphics[width=\linewidth]{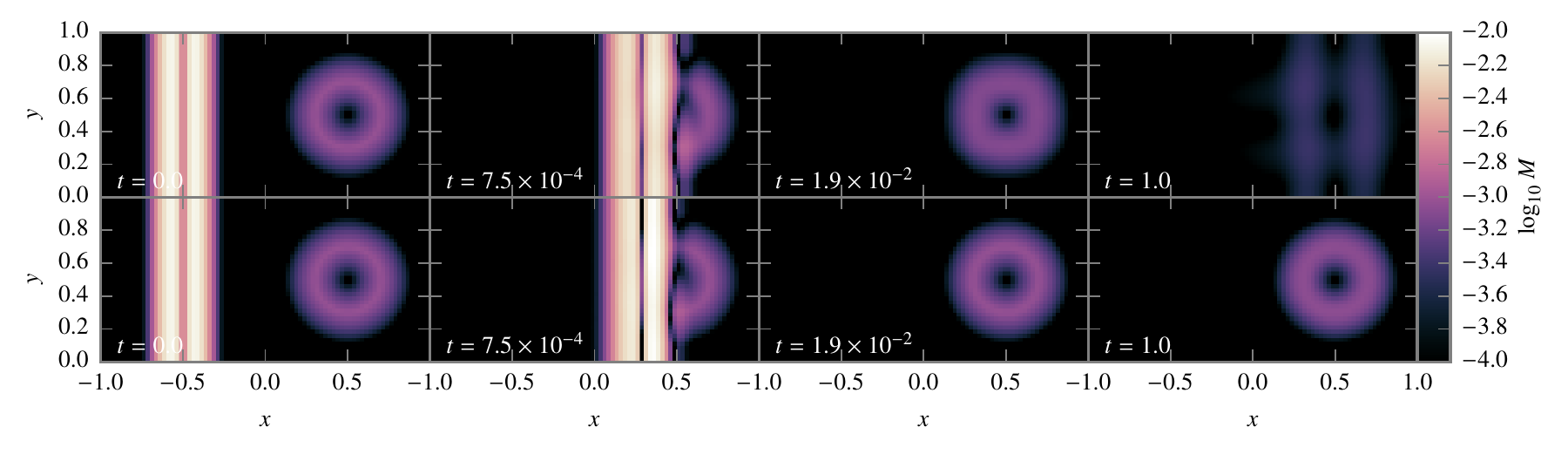}
\caption{Logarithmic Mach number of a plane sound wave packet
propagating through the Gresho vortex. The upper row shows the
unpreconditioned Roe solver, the lower row was computed using the
preconditioned Roe solver. The columns show the Mach number at the
beginning (left), when the wave is just passing through the vortex
(center left), some time after the sound wave has left the domain
(center right), and after a full rotation of the vortex (right).}
\label{fig:gresho_sound}
\end{figure*}

The initial Mach numbers of this setup are shown in the left panels of
Fig.~\ref{fig:gresho_sound}. We expect the wave to pass through the
vortex without interference as the compressible and incompressible
solutions decouple in the low-Mach limit (see the discussion in
Sect.~\ref{sect:lowmach}). The top and bottom boundary conditions of
the domain are periodic, as before. The left and right boundaries,
however, are now chosen to ensure that the wave leaves the domain
without producing numerical artifacts. This is achieved by
implementing the far-field boundary condition by
\citet{ghidaglia2005a}.

We computed the time evolution of the problem using the standard Roe
scheme. In a second run, the calculation was repeated, but this time we
applied our new preconditioning technique to the numerical fluxes. For
time stepping, we used the explicit RK3 method with the acoustic CFL
criterion because we need to limit the time steps to the sound crossing
time to resolve the sound wave. The Mach numbers for
different times are shown in Fig.~\ref{fig:gresho_sound}. The tests
demonstrate that the sound wave passes through the vortex without
interference in both runs, as expected. However, after one full
revolution, the high dissipation in the unpreconditioned case leads to
a destruction of the vortex (see upper right panel of
Fig.~\ref{fig:gresho_sound}). We emphasize that this is the same
effect as studied before and not caused by the interaction with the
sound wave that has left the domain long before. In contrast, the
vortex is preserved over one full revolution if preconditioning is
applied (lower right panel of Fig.~\ref{fig:gresho_sound}).

The test demonstrates that our preconditioning method, while
enabling to accurately represent flows at low Mach numbers, retains the
capability of modeling compressible hydrodynamics phenomena on the
same grid. The propagation of the sound waves in the unpreconditioned
and preconditioned cases is identical (compare upper and lower
panels in the second column of Fig.~\ref{fig:gresho_sound}). In
particular, its propagation speed is not altered by the preconditioning
technique. This shows that our method is perfectly capable of providing
the correct solution to problems that combine compressible and
incompressible hydrodynamical flows.

\section{Conclusions}
\label{sect:concl}

Many astrophysical problems involve hydrodynamical flows at low Mach
numbers $M$. Such slow flows are solutions of the Euler
equations. Based on an analysis of their solution space in the limit
$\mathit{M} \to 0$, we formulated two requirements for a numerical
scheme to correctly treat flows at low Mach numbers:
Requirement~\ref{constraint_1} is that for $\mathit{M} \to 0$
incompressible solutions should be retained, provided that the initial
condition was well-prepared. Requirement~\ref{constraint_2} states
that in the low-Mach-number limit the total kinetic energy should be
conserved.

Following the work of \citet{turkel1999a} and \citet{guillard2004a},
who discussed the failure of Godunov-type solvers to correctly reproduce slow
flows in numerical simulations, we argued that the reason for this
problem in the case of the Roe scheme is an inconsistent scaling of the
Roe matrix for $\mathit{M} \to 0$. The upwind viscosity term
arising from it can be seen as artificial -- introduced only to
stabilize the numerical scheme. It should not dominate the
physical flux in the low-Mach-number limit. To achieve consistency
between the scalings of the Roe matrix and the Jacobian of the
physical flux, preconditioning was proposed. We introduced a new
preconditioning matrix that ensures perfect consistency and
demonstrated the capability of the novel scheme to reproduce flows at low Mach
numbers on two examples: the Gresho vortex and the
Kelvin--Helmholtz instability.

The Gresho vortex is a solution of the incompressible Euler equations,
and we showed that our new scheme, in contrast to conventional
unpreconditioned approaches, reproduces it satisfactorily even at very
low Mach numbers. Thus, our method complies with
requirement~\ref{constraint_1}.

Of course, some kinetic energy is
lost as a result of discretization errors. This means that requirement~\ref{constraint_2} is not fully
satisfied. However, the loss of kinetic energy is substantially lower
than in conventional schemes, even at moderately low Mach numbers. At
very low Mach numbers the difference is even more pronounced. While
the motion of the Gresho vortex is dissipated away quickly in the
unpreconditioned scheme, we observed no deterioration in the quality of
the solution with lower Mach number in our approach, and it therefore
is applicable to almost arbitrarily slow flows. This is attributed to
the preconditioning of the upwind artificial viscosity. This term
dominates numerical dissipation at low Mach numbers in conventional
schemes, and this undesired effect is removed in our method.

We emphasize that our preconditioning technique only affects an
artificial term in the numerical flux that was introduced to
stabilize the scheme. Its choice is somewhat arbitrary. Thus, compared
to conventional methods, our new scheme is merely a different numerical
approximation to the solution of the same set of equations. This
distinguishes our technique from other approaches of numerically
 modeling flows
at low Mach
numbers. These usually start out from a modification
of the system to be solved.  Thus, our new method has the convenient
property that well-tested standard Godunov-type solvers can be used
(when properly preconditioned). In particular, we note that our
modifications do not affect the equation of state and there is no
restriction to an ideal gas. Well-known extensions of finite-volume
schemes to general equations of state can be used, which is important
for astrophysical applications.  In addition, in terms of the flow regimes to
be modeled, our scheme is more generally applicable. It can be used for
flows at low Mach numbers, but because the underlying equations are
unchanged, it also applies to the compressible regime, where
preconditioning is switched off by construction.

These special properties imply that our new scheme promises to
be able to successfully model hydrodynamic processes in stellar astrophysics that
are inaccessible for conventional methods. A thorough description of
its numerical implementation will be given in a forthcoming publication.

\begin{acknowledgements}
  We thank Konstantinos Kifonidis for pointing out Turkel's
  preconditioning technique to us. This inspired the development of
  the scheme presented here. We also gratefully acknowledge
  stimulating discussions with Christian Klingenberg and Markus Zenk.
  The work of FKR and PVFE is supported by the Deutsche
  Forschungs\-gemeinschaft (DFG) through the graduate school on
  ``Theoretical Astrophysics and Particle Physics'' (GRK 1147). FKR
  receives additional support from the Emmy Noether Program (RO 3676/1-1)
  of DFG and the ARCHES prize of the German Federal Ministry of
  Education and Research (BMBF).
\end{acknowledgements}

\appendix
\section{Variable transformation matrices}
\label{sect:var-trans}
Here we present the Jacobian matrices for the transformation between
conservative variables~$\fvec{U}$, defined in Eq.~\ref{eq:def-U}, and
primitive variables~$\fvec{V}$, defined in
Eq.~\ref{eq:def-V}. $\fvec{V}$ can be expressed in terms of the
components of $\fvec{U}$ by
\begin{equation*}
  \fvec{V} = 
  \left(
    \begin{array}{c}
      \fvec{U}_1 \\[1ex]
      \fvec{U}_2 / \fvec{U}_1 \\[1ex]
      \fvec{U}_3 / \fvec{U}_1 \\[1ex]
      \fvec{U}_4 / \fvec{U}_1 \\[1ex]
      p\left( \fvec{U}_1, \fvec{U}_5 - \frac{1}{2 \fvec{U}_1}
        \ndref{M}^2 
        \left( 
          \fvec{U}_2^2 + \fvec{U}_3^2 + \fvec{U}_4^2
        \right),
        \fvec{U}_6 / \fvec{U}_1
      \right) \\[1ex]
      \fvec{U}_6 / \fvec{U}_1
    \end{array}
  \right),
\end{equation*}
where $p(\rho, \epsilon, X)$ is an arbitrary equation of state,
depending on mass density, internal energy per volume~$\epsilon$, and composition.

The Jacobian matrix for the transformation from $\fvec{U}$ to $\fvec{V}$ can then be directly calculated:
\begin{equation*}
  \frac{\partial\fvec{V}}{\partial\fvec{U}} = 
  \left(
    \begin{array}{cccccc}
      1 & 0 & 0 & 0 & 0 & 0 \\[1ex]
      -\frac{u}{\rho} & \frac{1}{\rho} & 0 & 0 & 0 & 0\\[1ex]
      -\frac{v}{\rho} & 0 & \frac{1}{\rho} & 0 & 0 & 0\\[1ex]
      -\frac{w}{\rho} & 0 & 0 & \frac{1}{\rho} & 0 & 0\\[1ex]
      \alpha &
      -p_{\epsilon} \ndref{M}^2 u&
      -p_{\epsilon} \ndref{M}^2 v&
      -p_{\epsilon} \ndref{M}^2 w&
      p_{\epsilon} & \frac{p_{X}}{\rho} \\[1ex]
      -\frac{X}{\rho} & 0 & 0 & 0 & 0 & \frac{1}{\rho}
    \end{array}
  \right),
\end{equation*}
with
\begin{equation*}
  \alpha = p_{\rho} + \frac{1}{2} \ndref{M}^2 p_{\epsilon} 
      \left( 
        u^2 + v^2 + w^2
      \right) -  
      \frac{p_{X} X}{\rho}.
\end{equation*}
This introduces abbreviations for the derivatives of the equation of state,
\begin{equation*}
  p_{\rho} = \left.\frac{\partial p}{\partial\rho}\right|_{\epsilon,X}
  \qquad
  p_{\epsilon} = \left.\frac{\partial p}{\partial\epsilon}\right|_{\rho,X}
  \qquad
  p_{X} = \left.\frac{\partial p}{\partial X}\right|_{\rho,\epsilon}.
\end{equation*}

Inverting the above matrix yields the Jacobian for the conversion from $\fvec{V}$ to $\fvec{U}$,
\begin{equation*}
  \frac{\partial\fvec{U}}{\partial\fvec{V}}
  = 
  \left( \frac{\partial\fvec{V}}{\partial\fvec{U}} \right)^{-1}
  =
  \left(
    \begin{array}{cccccc}
      1 & 0 & 0 & 0 & 0 & 0 \\[1ex]
      u & \rho & 0 & 0 & 0 & 0 \\[1ex]
      v & 0 & \rho & 0 & 0 & 0 \\[1ex]
      w & 0 & 0 & \rho & 0 & 0 \\[1ex]
      \beta &
      \rho u \ndref{M}^2 &
      \rho v \ndref{M}^2 &
      \rho w \ndref{M}^2 &
      \frac{1}{p_{\epsilon}} &
      -\frac{p_{X}}{p_{\epsilon}} \\[1ex]
      X & 0 & 0 & 0 & 0 & \rho
    \end{array}
  \right),
\end{equation*}
with
\begin{equation*}
  \beta = \frac{1}{2} \ndref{M}^2
      \left( u^2 + v^2 + w^ 2 \right)
      - \frac{p_{\rho}}{p_{\epsilon}}.
\end{equation*}
An advantage of this approach is that it uses the same derivatives of the equation of state. This avoids differentiating the (possibly tabulated) equation of state in its inverted form.

\end{document}